\documentstyle[11pt,a4]{article}
\begin{document}
\newcommand{\nn}{\nonumber\\}
\vbox{\hbox{\hspace{12cm} DAMTP/96-27}
\title{\bf A New RSOS Restriction of the Zhiber-Mikhailov-Shabat Model
       and $\Phi_{(1,5)}$ Perturbations of Nonunitary 
       Minimal Models}
\author{ G\'abor Tak\'acs \\
         Department of Applied Mathematics and Theoretical Physics\\
         Cambridge, UK }
\date{15 April 1996}
\maketitle}
\begin{abstract}
The RSOS restriction of the Zhiber-Mikhailov-Shabat (ZMS) model is
investigated. It is shown that in addition to the usual RSOS
restriction, corresponding to $\Phi_{(1,2)}$ and $\Phi_{(2,1)}$
perturbations of minimal CFT, there is another one 
which yields $\Phi_{(1,5)}$ perturbations of non-unitary 
minimal models. The new RSOS restriction is carried out and the
particular case of the minimal models ${\cal M}_{(3,10)}$, ${\cal
M}_{(3,14)}$ and ${\cal M}_{(3,16)}$ is discussed in detail. In the
first two cases, while the mass spectra of the two RSOS restrictions
are the same, the bootstrap systems and the detailed amplitudes are
different. In the third case, even the spectra of the two RSOS
restrictions are different. In addition, for ${\cal M}_{(3,10)}$ an
interpretation in terms of the tensor product of two copies of ${\cal
M}_{(2,5)}$ is given.
\end{abstract}
\section{Introduction}

It is well-known that certain perturbations of minimal models of
conformal field theory (for a review see \cite{bpz}) lead to massive
integrable field theories \cite{pertcft}. In particular, the
perturbations described by the operators  
$\Phi_{(1,2)},\ \Phi_{(1,3)},\ \Phi_{(2,1)}$ 
have this property. The description of these massive theories is given
in terms of certain restrictions of other integrable field
theories. Namely, the $\Phi_{(1,3)}$ perturbations are described 
as RSOS restrictions of sine-Gordon theory \cite{smirresh,berlecl},
while the other two cases can be treated in terms of the
Zhiber-Mikhailov-Shabat (ZMS) model \cite{smirnov,efthimiou}.

The sine-Gordon and ZMS models have quantum affine symmetry algebras 
${\cal U}_{q}(A_1^{(1)})$ \cite{sgqaff} and ${\cal U}_{q}(A_2^{(2)})$
\cite{efthimiou}, respectively. These algebras are generated by
non-local conserved charges. The invariance under these quantum
symmetries determines the S-matrix up to a scalar factor
\cite{smirnov,efthimiou,sgqaff}. If the value of $q$ is a root of
unity, then the representation theory of the quantum group allows a 
consistent truncation to a maximal spin, which is known as the RSOS
restriction. In the case of sine-Gordon theory, the algebra ${\cal
U}_{q}(A_1^{(1)})$ has two copies of ${\cal U}_{q}(sl(2))$ as
subalgebras, both of which can be chosen to perform the restriction
\cite{smirresh}.  These two subalgebras give the same reduced theory
for the reason that there is a simple automorphism of the quantum
affine symmetry algebra interchanging the two subalgebras. 

However, this not true for the ZMS model. ${\cal U}_{q}(A_2^{(2)})$
has the two subalgebras ${\cal U}_{q}(sl(2))$ and 
${\cal U}_{q^4}(sl(2))$, both of which can be chosen for the
restriction. The choice leading to the $\Phi_{(1,2)}$ and
$\Phi_{(2,1)}$ perturbations is the first one, and this is the idea
pursued in \cite{smirnov,efthimiou}. In that case the fundamental
particles of ZMS model form a triplet representation under the first
subalgebra, which is irreducible. Under the second subalgebra the
triplet decomposes as a doublet and a singlet, leaving one with the
task of disentangling the amplitudes mixing these two
components. Hence the restriction procedure turns out to be more
complicated, but as will be shown in the sequel, it can be analysed
and made systematic. 

The goal of this paper is to examine the second possibility and to
obtain the new RSOS restriction of the ZMS model which emerges from
it. It will be argued that the S-matrices obtained in this way
correspond to $\Phi_{(1,5)}$ perturbations of minimal models. 
The S-matrix for the case ${\cal M}_{(2,9)}$, which is a particular 
example, 
has been obtained by Martins in \cite{martins1}, where he also raised the 
question whether it is possible to generalise the RSOS restriction to 
the $\Phi_{(1,5)}$ perturbations. In the framework presented here  
this goal can be achieved in generality. In particular, Martins' 
S-matrix is reconstructed as a special case in subsection 5.1.

It is also natural to expect that there exist inequivalent RSOS restrictions 
for other imaginary coupling Toda field theories based on 
non-simply laced affine Kac-Moody algebras, too. It may be worthwhile to
undertake the investigation of these theories in the future.

This article is part of a larger project devoted to the study of 
$\Phi_{(1,5)}$ perturbations of minimal models and their connections
to $\Phi_{(1,2)}$ perturbations through the ZMS model. The other part
of the work has also been completed and the paper is currently in
preparation \cite{our}. It will contain detailed analysis of the
models treated in Section 5, using the method of the thermodynamical
Bethe Ansatz (TBA) \cite{tba1,tba2,tba3}
and the truncated conformal space approach (TCSA) \cite{tcsa1,tcsa2}.

The layout of the paper is the following. Section 2 gives an
introduction and brief review of the S-matrix of the ZMS model, mainly
in order to set up notations. Section 3 is devoted to the definition
and examples of related perturbed minimal models, 
which are the two possible restrictions of
the same ZMS model. Section 4 describes the general strategy
of RSOS restriction, taking sine-Gordon theory as an example, 
which is then applied to the ZMS model to find the new RSOS
restriction. Section 5 contains the detailed discussion of some
examples. The focus will be on the $\Phi_{(1,5)}$ perturbations of the
theories ${\cal M}_{(3,10)}$, ${\cal M}_{(3,14)}$ and 
${\cal M}_{(3,16)}$, since these are the ones treated in \cite{our}. 
Section 6 is reserved for the conclusions. 
\section{Review of the S-matrix of the ZMS model}
\label{sect2} 

The ZMS model is defined by the Lagrangian
\begin{equation}
{\cal L}=\int \left((\partial_\mu\phi)^2+\frac{m^2}{\gamma^2}
\left(\exp(i\sqrt{8\gamma}\phi)
+\exp(-i\sqrt{2\gamma}\phi)\right)\right)dx\ ,
\label{zmslagr}\end{equation}
with $m$ being a mass parameter and $\gamma$ is the coupling
constant. 

The model given by (\ref{zmslagr}) is an imaginary coupling 
affine Toda theory based on the twisted affine 
Kac-Moody algebra $A_2^{(2)}$. This algebra is non-simply laced with
two roots, corresponding to the two exponential terms in the 
potential. The Hamiltonian of the model is not hermitian and therefore
the model is not unitary. This is in essentially different from
sine-Gordon theory, which corresponds to the untwisted affine
Kac-Moody algebra $A_1^{(1)}$ and has a hermitian Hamiltonian. 

However, as has been shown by Smirnov \cite{smirnov}, the model is
reducible for special values of the coupling $\gamma$ and the
restrictions correspond to perturbations of minimal models, among them
to unitary ones. More specifically, the model with $\gamma=\pi (r/s)$
can be reduced to the $\Phi_{(1,2)}$ perturbation of the minimal model
${\cal M}_{r,s}$ with central charge 
\begin{equation}
c=1-\frac{6(r-s)^2}{rs}\ .
\end{equation}
Therefore, while quantum field theory interpretation interpretation of 
the original ZMS model does not seem to be straightforward, due to its 
nonunitarity, we can approach the problem by taking the Lagrangian 
(\ref{zmslagr}) to describe a putative model which can be given at 
least after carefully restricting it to some Hilbert space. On the other 
hand, there are statistical systems (e.g. the Yang-Lee edge singularity) 
whose description leads to nonunitary models of two-dimensional field 
theory and so we can hope that the models derived from (\ref{zmslagr}) 
make sense in the realm of two-dimensional critical phenomena.

In this section I first outline the derivation of the ZMS S-matrix
following \cite{efthimiou}, in order to set up conventions and fix
some typos in the formulae given in \cite{smirnov,efthimiou}.
Then I briefly discuss the restriction carried out in \cite{smirnov}.

\subsection{The quantum symmetry of the ZMS model}

Using the Lagrangian (\ref{zmslagr}) it is possible to construct
non-local charges commuting with the Hamiltonian. In \cite{efthimiou}
it is shown that the nonlocal charges generate the quantum affine
algebra ${\cal A}={\cal U}_{q}(A_2^{(2)})$ with 
\begin{equation}
q=\exp (i\pi^2/\gamma )\ . 
\end{equation}
The defining relations of this algebra are
\begin{eqnarray}
&&[H_0,H_1]=0 ,\nn
&&[H_i,E_j]=+a_{ij}E_j ,\nn
&&[H_i,F_j]=-a_{ij}F_j ,\nn 
&&[E_i,F_j]=\delta_{ij}\frac{q^{H_i}-q^{-H_i}}{q_i-q_i^{-1}} ,
\end{eqnarray}
where $a_{ij}$ is the symmetrized Cartan matrix of $A_2^{(2)}$
\begin{equation}
[a_{ij}]=\left[\matrix{ 8 & -4 \cr -4 & 2 }\right]
\end{equation}
and 
\begin{equation}
q_i=q^{a_{ii}/2}
\end{equation}
The fundamental representation of $\cal A$ is three-dimensional
and is given by the following matrices:
\begin{eqnarray}
&&H_0 = \left [\begin {array}{ccc} -4&0&0\\ \noalign{\medskip}0&0&0
\\ \noalign{\medskip}0&0&4\end {array}\right ] ,\nn
&&E_0 = \left [\begin {array}{ccc} 0&0&0\\ \noalign{\medskip}0&0&0
\\ \noalign{\medskip}1&0&0\end {array}\right ] ,\quad
F_0 = \left [\begin {array}{ccc} 0&0&1\nn\noalign{\medskip}0&0&0
\\ \noalign{\medskip}0&0&0\end {array}\right ] ,\nn
&&H_1 = \left [\begin {array}{ccc} 2&0&0\\ \noalign{\medskip}0&0&0
\\ \noalign{\medskip}0&0&-2\end {array}\right ] ,\nn
&&E_1 = \left [\begin {array}{ccc} 0&1&0\\ \noalign{\medskip}0&0&-1
\\ \noalign{\medskip}0&0&0\end {array}\right ] ,\quad
F_1 = \left [\begin {array}{ccc} 0&0&0\\ \noalign{\medskip}
\left (q+{q}^{-1}
\right )&0&0\\ \noalign{\medskip}0&-\left (q+{q}^{-1}\right )&0
\end {array}\right ]\ .
\label{fundrep}\end{eqnarray}
The coproduct of the algebra $\cal A$ is 
\begin{eqnarray}
&&\Delta (H_i)=H_i\otimes 1+1 \otimes H_i,\nn
&&\Delta (E_i)=E_i\otimes q^{-H_i/2}+q^{H_i/2} \otimes E_i,\nn
&&\Delta (F_i)=F_i\otimes q^{-H_i/2}+q^{H_i/2} \otimes F_i,
\label{coproduct}\end{eqnarray}
and can be verified to be an algebra homomorphism ${\cal A}\rightarrow
{\cal A} \otimes {\cal A}$. It is possible to define a counit and an 
antipode, but they will not be needed here.

It is important to observe that there are two copies of ${\cal
U}_p(sl(2))$ inside the algebra $\cal A$: the generators 
$\{ H_1,\ E_1,\ F_1\}$ form ${\cal U}_q(sl(2))$, while $\{ H_0,\ E_0,\
F_0\}$ form ${\cal U}_{q'}(sl(2))$, with $q'=q^4$. 
These subalgebras will be denoted by ${\cal A}_1$ and ${\cal A}_0$, 
respectively.

The fundamental particles of the model are a triplet of kinks 
and the corresponding one-particle states form 
a  so-called evaluation representation of $\cal A$, 
which is a rapidity-parametrized version of the fundamental
representation described in (\ref{fundrep}). More precisely,
we can introduce the triplet of asymptotic one-particle states as
\begin{equation}
| \theta, i \rangle_{as} ,
\end{equation}
where $\theta$ is the usual rapidity variable connected to the
energy-momentum as $p^0=m \cosh\theta\ ,\ p^1=m \sinh\theta$, with $m$
being the particle mass and $i$ labels the three components. The
eigenvalues of $H_1$ on these states give the topological charge of
the kink. The label ``as'' can assume the values ``in'' and
``out''. The general multiparticle in-states and out-states are
\begin{eqnarray}
&&|\theta_1, i_1, \theta_2, i_2, \dots \theta_n, i_n \rangle_{in}\nn  
&&|\theta_n, i_n, \dots \theta_2, i_2, \theta_1, i_1 \rangle_{out} ,
\end{eqnarray}
with the rapidities ordered as $\theta_1 > \dots > \theta_n$.
The evaluation representation $Ev_{\theta}$ is defined by
\begin{equation}
e_i = x_i E_i\ ,\ f_i=x_i^{-1}F_i\ ,\  x_i=\exp(s_i\theta ),
\end{equation}
with $s_0=4\pi/\gamma -1\ ,\ s_1=\pi/\gamma -1$
(cf. \cite{efthimiou}). This particular choice of the rapidity
dependence corresponds to the so-called spin gradation. The action on
the multiparticle states is easily derived using the coproduct
(\ref{coproduct}). For details on quantum affine algebras and their
representation theory cf. \cite{chari}.

\subsection{The S-matrix}

The invariance of the S-matrix can be formulated as follows.
The two-particle S-matrix maps the in-states to the out-states:
\begin{equation}
|\theta_2, j_2, \theta_1, j_1 \rangle_{out}=
S^{j_1j_2}_{i_1i_2}(\theta_1-\theta_2)
|\theta_1, i_1, \theta_2, i_2,\rangle_{in}\nn  
\end{equation}
We require that the two-particle S-matrix must be invariant  
under the action of the quantum affine symmetry algebra on the 
asymptotic states. Let us define the matrix 
\begin{equation}
R(x,q)=P_{12}{\hat S}\ ,
\end{equation} 
where $P_{12}$ is the permutation operator
acting on the internal indices of the two-particle state and 
${\hat S}$ denotes the tensor part of the S-matrix. The quantum symmetry 
specifies the S-matrix only up to a scalar function - the precise 
normalization of the tensor part will be clear from (\ref{hatS}) 
below. We also change the gradation to the so-called homogeneous one, 
in which the complete rapidity dependence is carried by $E_0$ and $F_0$:
\begin{equation}
e_0^h=xE_0\ ,\ f_0^h=x^{-1}F_0\ ,\ e_1^h=E_1\ ,\ f_1^h=F_1\ ,\
x=x_0x_1^2\ . 
\end{equation}
This can be implemented by the following algebra automorphism:
\begin{equation}
{\cal A}\rightarrow {\cal A}\ ,\ a\mapsto x_1^{-H_1/2}ax_1^{H_1/2}.
\label{chgrad}\end{equation}
The invariance of the S-matrix implies that $R(x,q)$ has to intertwine
between the representations $Ev_{\theta_1}\otimes Ev_{\theta_2}$ and 
$Ev_{\theta_2}\otimes Ev_{\theta_1}$. With the variable $x$ defined in
terms of the relative rapidity $\theta=\theta_1-\theta_2$, $R(x,q)$ is
given by  
\begin{equation}
R(x,q)=\frac{(1-q^4)(q^6+1)}{q^5}P_{12}+\frac{x-1}{q^3}R_{12}+
\frac{q^3(x-1)}{x}R_{21}^{-1},
\label{hatS}\end{equation}
where 
\begin{equation}
R_{12}=\left [\matrix{ {q}^{-2}&0&0&0&0&0&0&0&0
\cr 0&1&0&-{\frac {{q}^{4}-1}{{q}^{2}}}&0&0&0&0&0
\cr 0&0&{q}^{2}&0&{\frac {{q}^{4}-1}{q}}&0&{\frac {-{q
}^{4}+{q}^{6}-{q}^{2}+1}{{q}^{2}}}&0&0\cr 0&0&0&1&0&0&0
&0&0\cr 0&0&0&0&1&0&{\frac {{q}^{4}-1}{q}}&0&0
\cr 0&0&0&0&0&1&0&-{\frac {{q}^{4}-1}{{q}^{2}}}&0
\cr 0&0&0&0&0&0&{q}^{2}&0&0\cr 0&0&0&0
&0&0&0&1&0\cr 0&0&0&0&0&0&0&0&{q}^{-2}}
\right ] . 
\end{equation}
It plays a
central role in the investigation of the ZMS model in the framework of
quantum inverse method and was originally obtained by Izergin and
Korepin in that context \cite{izergin}.

$R(x,q)$ is a solution of the parametric quantum Yang-Baxter equation:
\begin{equation}
R_{12}(x/y,q)R_{13}(x,q)R_{23}(y,q)=R_{23}(y,q)R_{13}(x,q)R_{12}(x/y,q)
\ .
\end{equation}
and satisfies the following important relations, written in terms of 
${\hat S}(x)$:
\begin{eqnarray}
&&{\hat S}(x){\hat S}(1/x)=
{\frac {\left (x+{q}^{6}\right )\left (x-{q}^{4}\right
)\left (1+{q}^{6}x\right )\left (1-x{q}^{4}\right
)}{{q}^{10}{x}^{2}}},\nn
&&{\hat S}(-q^6/x)=
(1\otimes \alpha ){\hat S}_{21}^{t_1}(x,q)(1\otimes \alpha ),
\label{crossuni}\end{eqnarray}
with 
\begin{equation}
\alpha = \left [\matrix{ 0&0&{q}^{-1}\cr 0&1&0 \cr q&0&0}\right ]\ .
\end{equation}
The first property in (\ref{crossuni}) ensures ``unitarity'', while
the second means that in the spin gradation the S-matrix is actually
crossing symmetric. The term ``unitarity'' must be understood
carefully: here we refer to the property of the S-matrix given by 
\begin{equation}
S_{ij}^{kl}(-\theta)S_{kl}^{mn}(\theta )=\delta_i^m\delta_j^n\ ,
\label{unitarity}\end{equation}
while the quantum field theory itself is not unitary, which can be
seen e.g. from the negative sign of the residues at the poles of the
S-matrices of singlet bound states in the theory.

The solution for the S-matrix is fixed by its intertwining property 
up to multiplication with a scalar function. To make it unitary, while 
preserving crossing symmetry, it has to be multiplied by the following 
factor \cite{smirnov} (with an overall sign that can
be chosen arbitrarily):
\begin{eqnarray}
S_0(\theta )=
\pm\frac{1}{4i}\left( \sinh\frac{\pi}{\xi}\left(\theta -\pi i\right)
\sinh\frac{\pi}{\xi}
\left(\theta -\frac{2\pi i}{3}\right)\right)^{-1}\times\nn
\exp\left( -2i\int\limits_0^\infty 
\frac{\sin k\theta \sinh \frac{\pi k}{3} 
\cosh\left(\frac{\pi}{6}-\frac{\xi}{2}\right) k}
{k\cosh\frac{\pi k}{2}\sinh\frac{\xi k}{2} } dk \right)\ ,
\label{S0}\end{eqnarray}
with $\xi$ given by
\begin{equation}
\xi=\frac{2}{3}\left(\frac{\pi\gamma}{2\pi -\gamma}\right)\ .
\end{equation}
Using $\xi$, the variable $x$ can be alternatively written as 
\begin{equation}
x=\exp\left(\frac{2\pi\theta}{\xi}\right) .
\end{equation}
In the forthcoming paper \cite{our}, where the S-matrices proposed
later in this work are put to some stringent tests, 
it will be shown that the sign choice in (\ref{S0}) is tied up with the
statistics of the particle and does not have any physical meaning in
itself (in two dimensions, statistics has no inherent meaning either,
only in conjunction with the choice of whether $S(\theta =0)$ is $+1$
or $-1$, see in the context of the thermodynamical Bethe Ansatz, e.g. 
\cite{tba1}).

The S-matrix given above is just the S-matrix of the lowest lying
state, which is the fundamental kink triplet. If one needs to have the
complete S-matrix, it can be completed using the usual methods 
of the S-matrix bootstrap. In this way, generically, higher kinks
and breathers arise. The fundamental kink S-matrix has poles at
$\theta =i\pi - i\xi m$ and $\theta = 2i\pi /3 - i\xi m$, with $m$
being an integer. The ones in the physical strip
$0\leq \Im m \theta < \pi$ correspond to bound states. The first
sequence corresponds to breathers. At these points ${\hat S}(x)$
degenerates into a rank-one projector, indicating that there is a
singlet bound state. The second sequence corresponds to higher kinks,
since there ${\hat S}(x)$ degenerates into a rank-three projector,
corresponding to a triplet of particles. All poles have their crossing
symmetric counterparts at $\theta = i\xi m$ and 
$\theta = i\pi /3 + i\xi m$, respectively. The breather-kink and
breather-breather S-matrices are all scalars, while the kink-kink
S-matrices have the following general form:
\begin{equation}
{\hat S}\left(\exp\left(\frac{2\pi}{\xi}\theta + 
i\phi(k_1,k_2)\right)\right) S^{k_1k_2}_0(\theta ),
\label{higherkinkS}\end{equation}
as shown in \cite{smirnov}. Here $k_1,\ k_2$ number the kinks,
$S^{k_1k_2}_0(\theta )$ is a scalar function similar to $S_0(\theta )$
and $\phi(k_1,k_2)$ is a phase shift satisfying
$\phi(k_1,k_2)=-\phi(k_2,k_1)$.
I wish to note, however, that the formula given in \cite{smirnov} for
the explicit expression of the tensor part of the kink-higher kink
S-matrix (eqn. (3.7) in \cite{smirnov}) is off by a scalar factor.  
The correct version is
\begin{equation}
A{\cal P}_{12}R_{13}(xq^2,q)R_{23}(x/q^2,q){\cal P}_{12}A^{-1}
=\frac{(1-xq^2)(x+q^8)}{q^5x}R_{(12)3}(-x,q),
\label{hkfusion}\end{equation}
where ${\cal P}_{12}$ is the projector on the spin-$1$ representation 
in the product of two spin-$1$ representations, the index $(12)$ 
denotes its $3$-dimensional image,
and $A$ denotes a redefinition of the states given by the matrix 
\begin{equation}
\left[\matrix{I_3 & 0 & 0 \cr
0 &-{\frac {-4\,{q}^{2}+{q}^{4}+1}{2\,{q}^{2}}}I_3 &0 \cr  
0 & 0 & I_3}\right]\ 
\end{equation}
(with $I_3$ representing the $3\times 3$ unit matrix), 
which is equivalent to redefining the middle component of the higher 
kink triplet.

Note that the fundamental distinction between a kink and a breather is
given by the fact that kinks are not singlets under the quantum
algebra, they come in triplets, however, breathers are singlets. This
essentially amounts to saying that while kinks carry topological
charge, the breathers do not, since the topological charge is part of
the quantum algebra.

\subsection{RSOS restriction of ZMS model and $\Phi_{(1,2)}$
perturbations of minimal models}

The RSOS restriction, described in \cite{smirnov,efthimiou}, proceeds
in the following way. If $q$ is a root of unity, then we know that the
representation of the quantum algebra is different from the case of
generic $q$. Supposing that 
\begin{equation}
\gamma=\pi (r/s)\ ,
\end{equation}
we get $q^r=\pm 1$. In
this case it is possible to consistently truncate the Hilbert space to
representations of ${\cal A}_1$, which have spins not exceeding 
$j_{max}=(r-2)/2$. The details of the construction are given in numerous
papers in the context of sine-Gordon theory (see
e.g. \cite{smirresh,berlecl}) as well as the ZMS model
(e.g. \cite{smirnov,efthimiou}) and it will be outlined in the section
dealing with the other RSOS restriction of the ZMS model.

The truncation of tensor product is the same which occurs in minimal
models of conformal field theory (see \cite{bpz}; for a detailed
exposition of the quantum group structure of minimal models, see
\cite{gomez} and references therein). Indeed, the models obtained in
this way can be considered as perturbations of minimal models 
${\cal M}_{r,s}$ with the field $\Phi_{(1,2)}$ in the Kac table. This
is known to be a relevant and integrable perturbation of all minimal
models \cite{pertcft}. 

The way, in which this identification arises, is crucial to the
considerations of this paper. As shown in \cite{liouville1,liouville2}
the minimal model ${\cal M}_{r,s}$ can be considered as the conformal
quantization of the imaginary coupling Liouville field theory given by
\begin{equation}
{\cal L}=\int((\partial_\mu\phi)^2+\exp(i\sqrt{8\gamma}\phi))dx\ .
\end{equation}
The operators 
\begin{equation}
\exp\left(-i\frac{n-1}{2}\sqrt{8\gamma}\phi\right)
\end{equation}
can be identified with the primary fields $\Phi_{(1,n)}$. It is then
easy to see the correspondence between the ZMS model at the coupling 
$\gamma=\pi (r/s)$ and the $\Phi_{(1,2)}$ perturbation of the minimal
model ${\cal M}_{r,s}$.
\section{Relation to $\Phi_{(1,5)}$ perturbations}\label{sect3}

The identification of ZMS model and perturbed minimal models,
described at the end of the previous section, lends itself to the
following idea. Interchanging the role of the exponentials in
(\ref{zmslagr}), 
i.e. taking the second one as the part of the imaginary coupling
Liouville model and the first one as perturbation, one arrives at
another integrable model, which on the level of the ZMS model is the
same as the starting point. The question is: what happens after RSOS
restriction? 

In sine-Gordon theory, where the two exponentials have the same
exponents with opposite signs, this interchange of the two terms can
be compensated by a simple automorphism of the quantum symmetry
algebra. Here this is not the case, because the two roots have
different lengths. Generically, such reshuffling is expected to yield
different models after RSOS restriction for any imaginary coupling
Toda field theory associated to a non-simply laced quantum affine
algebra. 

It can be easily checked, using the imaginary coupling Liouville
theory description, that the new model will correspond to 
a $\Phi_{(1,5)}$ perturbation of another minimal model 
${\cal M}_{r',s'}$ with
\begin{equation}
\frac{r'}{s'}=\frac{1}{4}\frac{r}{s}\ . 
\end{equation}
The $\Phi_{(1,5)}$ perturbation of a unitary minimal model is
irrelevant. Hence we can expect only non-unitary cases to be
interesting. Some examples:
\begin{itemize}
\item The model ${\cal M}_{8,9}+\Phi_{(1,2)}$ is related to the model 
${\cal M}_{2,9}+\Phi_{(1,5)}$, which is the same (due to the symmetry 
of the Kac table) as the model ${\cal M}_{2,9}+\Phi_{(1,4)}$. This 
correspondence was used to prove the integrability and to conjecture
the S-matrix of this particular case by Martins et al. 
\cite{martins1,martins2}.
\item The case of the magnetic perturbation of the Ising model, i.e.
${\cal M}_{3,4}$. It is known to have an $E_8$ S-matrix. The
related model is the $\Phi_{(1,5)}$ perturbation of 
${\cal M}_{3,16}$.
\item The $\Phi_{(2,1)}$ perturbation of ${\cal M}_{5,6}$, which can
be thought of as the $\Phi_{(1,2)}$ perturbation of ${\cal
M}_{6,5}$. It is related by the above correspondence to the 
$\Phi_{(1,5)}$ perturbation of ${\cal M}_{3,10}$.
\item The $\Phi_{(1,2)}$ perturbation of ${\cal M}_{6,7}$ gets mapped
into the $\Phi_{(1,5)}$ perturbation of ${\cal M}_{3,14}$.
\end{itemize}

In this paper the focus will be on the models ${\cal M}_{3,10}$,
${\cal M}_{3,14}$ and ${\cal M}_{3,16}$. These are the models which
are subjected to the TCSA and TBA analysis in \cite{our}. 

\section{The RSOS restriction of the ZMS model}\label{sect4}

\subsection{RSOS restriction}

First I briefly recall some necessary facts about RSOS restriction, 
which will prove useful later in the study of the ZMS case. A more detailed 
exposition can be found in e.g. \cite{berlecl}.Take a doublet of solitons, 
transforming under ${\cal U}_q(sl(2))$. The tensorial part of the S-matrix 
will be a linear combination of the following matrix  
\begin{equation}
{\hat {\cal R}}=\sqrt{q}P_{12}{\cal R}^{\frac{1}{2}\frac{1}{2}}=
\left[ \matrix{ q&0&0&0 \cr 0&q-1/q&1&0 \cr 0&1&0&0 \cr 0&0&0&q
}\right] 
\label{doubletR}\end{equation}
and its inverse.

The many-particle Hilbert space can be decomposed into irreducible
representations of ${\cal U}_q(sl(2))$:
\begin{equation}
{\cal V}_{\frac{1}{2}}^{\otimes
N}=\mathop{\oplus}\limits_{\frac{1}{2},j_2,\dots j_N} {\cal V}_{j_N}\
,
\end{equation}
where ${\cal V}_{j}$ is the spin-$j$ representation of the quantum
group and $j_{i+1}=j_i\pm \frac{1}{2}\geq 0$ are the intermediate
representations in the $N$-fold tensor product. It can be represented
graphically as 

\setlength{\unitlength}{0.012500in}%
\begingroup\makeatletter\ifx\SetFigFont\undefined
\def\x#1#2#3#4#5#6#7\relax{\def\x{#1#2#3#4#5#6}}%
\expandafter\x\fmtname xxxxxx\relax \def\y{splain}%
\ifx\x\y   
\gdef\SetFigFont#1#2#3{%
  \ifnum #1<17\tiny\else \ifnum #1<20\small\else
  \ifnum #1<24\normalsize\else \ifnum #1<29\large\else
  \ifnum #1<34\Large\else \ifnum #1<41\LARGE\else
     \huge\fi\fi\fi\fi\fi\fi
  \csname #3\endcsname}%
\else
\gdef\SetFigFont#1#2#3{\begingroup
  \count@#1\relax \ifnum 25<\count@\count@25\fi
  \def\x{\endgroup\@setsize\SetFigFont{#2pt}}%
  \expandafter\x
    \csname \romannumeral\the\count@ pt\expandafter\endcsname
    \csname @\romannumeral\the\count@ pt\endcsname
  \csname #3\endcsname}%
\fi
\fi\endgroup
\begin{equation}
\begin{picture}(300,85)(30,740)
\thinlines
\put( 80,760){\line( 0, 1){ 40}}
\put(120,760){\line( 0, 1){ 40}}
\put(240,760){\line( 0, 1){ 40}}
\put(280,760){\line( 0, 1){ 40}}
\put(160,775){\makebox(0,0)[lb]{\smash{\SetFigFont{20}{24.0}{rm}   ...}}}
\put( 60,780){\makebox(0,0)[lb]{\smash{\SetFigFont{10}{14.4}{it}1/2}}}
\put(100,780){\makebox(0,0)[lb]{\smash{\SetFigFont{10}{14.4}{it}1/2}}}
\put(220,780){\makebox(0,0)[lb]{\smash{\SetFigFont{10}{14.4}{it}1/2}}}
\put(260,780){\makebox(0,0)[lb]{\smash{\SetFigFont{10}{14.4}{it}1/2}}}
\put( 75,810){\makebox(0,0)[lb]{\smash{\SetFigFont{12}{14.4}{it}a}}}
\put(115,810){\makebox(0,0)[lb]{\smash{\SetFigFont{12}{14.4}{it}a}}}
\put( 40,760){\line( 1, 0){280}}
\put(235,810){\makebox(0,0)[lb]{\smash{\SetFigFont{12}{14.4}{it}a}}}
\put(260,740){\makebox(0,0)[lb]{\smash{\SetFigFont{10}{12.0}{it}N-1}}}
\put(275,810){\makebox(0,0)[lb]{\smash{\SetFigFont{12}{14.4}{it}a}}}
\put( 30,750){\makebox(0,0)[lb]{\smash{\SetFigFont{12}{14.4}{it}0}}}
\put(325,750){\makebox(0,0)[lb]{\smash{\SetFigFont{12}{14.4}{it}j}}}
\put( 95,745){\makebox(0,0)[lb]{\smash{\SetFigFont{12}{14.4}{it}j}}}
\put(255,745){\makebox(0,0)[lb]{\smash{\SetFigFont{12}{14.4}{it}j}}}
\put( 80,805){\makebox(0,0)[lb]{\smash{\SetFigFont{10}{12.0}{it}1}}}
\put(120,805){\makebox(0,0)[lb]{\smash{\SetFigFont{10}{12.0}{it}2}}}
\put(240,805){\makebox(0,0)[lb]{\smash{\SetFigFont{10}{12.0}{it}N-1}}}
\put(280,805){\makebox(0,0)[lb]{\smash{\SetFigFont{10}{12.0}{it}N}}}
\put(100,740){\makebox(0,0)[lb]{\smash{\SetFigFont{10}{12.0}{it}1}}}
\put(330,745){\makebox(0,0)[lb]{\smash{\SetFigFont{10}{12.0}{it}N}}}
\end{picture}
\end{equation}

In this way we introduce a new labelling of the multiparticle Hilbert
space. Instead of identifying the states by the topological charge,
which is the eigenvalue of $H_0$, we decompose the space with respect
to the intermediate representation of the tensor product.

The RSOS restriction amounts to truncating the Hilbert space up to 
the spin $j_{max}=p/2-1$, where $p$ is the first positive integer for
which $q^p=\pm 1$. It corresponds to the truncation of the algebra
${\cal U}_q(sl(2))$ to ${\cal U}^{res}_q(sl(2))$ in the notations
\cite{chari}. In terms of the representation labels, there are only 
finitely many sectors left in the Hilbert space. The sine-Gordon
S-matrix maps the truncated space back unto itself, so it is
consistent to take the restriction.

Some care must be taken about the Hilbert space structure in the
restricted space. The original theory has a Hilbert space of its
own, but the restriction eliminates part of it. Hence it is
necessary to introduce a new inner product on the restricted space. 
The new product, however, turns out to 
define a unitary theory only in special cases, when a Hermitian
structure can be found. That is possible e.g. in the examples 
corresponding to perturbations of 
the unitary series of minimal models. This is described
in \cite{smirresh}, to which the interested reader is referred for 
more details. In the ZMS case, the model is not unitary to start
with, however, after the RSOS restriction, it is still possible to end
up with unitary theories \cite{smirnov}.

The doublet solitons yield kinks going back and forth between the
sectors. A kink $K_{ab}$ is labelled 
by the two representations $a$ and $b$ between which it
interpolates. The possible scattering processes look like 
\begin{equation}
K_{ab}(\theta_1 )+K_{bc}(\theta_2 ) \rightarrow 
K_{ad}(\theta_2 )+K_{dc}(\theta_1 )
\end{equation}
In graphical terms this process is conventionally represented as 
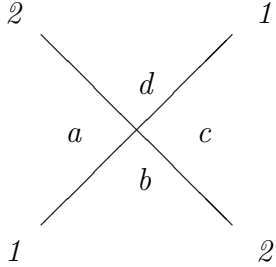
\setlength{\unitlength}{0.012500in}%
\begingroup\makeatletter\ifx\SetFigFont\undefined
\def\x#1#2#3#4#5#6#7\relax{\def\x{#1#2#3#4#5#6}}%
\expandafter\x\fmtname xxxxxx\relax \def\y{splain}%
\ifx\x\y   
\gdef\SetFigFont#1#2#3{%
  \ifnum #1<17\tiny\else \ifnum #1<20\small\else
  \ifnum #1<24\normalsize\else \ifnum #1<29\large\else
  \ifnum #1<34\Large\else \ifnum #1<41\LARGE\else
     \huge\fi\fi\fi\fi\fi\fi
  \csname #3\endcsname}%
\else
\gdef\SetFigFont#1#2#3{\begingroup
  \count@#1\relax \ifnum 25<\count@\count@25\fi
  \def\x{\endgroup\@setsize\SetFigFont{#2pt}}%
  \expandafter\x
    \csname \romannumeral\the\count@ pt\expandafter\endcsname
    \csname @\romannumeral\the\count@ pt\endcsname
  \csname #3\endcsname}%
\fi
\fi\endgroup
\begin{center}
\begin{picture}(105,115)(25,725)
\thinlines
\put( 40,740){\line( 1, 1){ 80}}
\put( 50,775){\makebox(0,0)[lb]{\smash{\SetFigFont{12}{14.4}{it}a}}}
\put( 80,755){\makebox(0,0)[lb]{\smash{\SetFigFont{12}{14.4}{it}b}}}
\put(105,775){\makebox(0,0)[lb]{\smash{\SetFigFont{12}{14.4}{it}c}}}
\put( 40,820){\line( 1,-1){ 80}}
\put( 80,795){\makebox(0,0)[lb]{\smash{\SetFigFont{12}{14.4}{it}d}}}
\put(130,825){\makebox(0,0)[lb]{\smash{\SetFigFont{12}{14.4}{it}1}}}
\put( 25,725){\makebox(0,0)[lb]{\smash{\SetFigFont{12}{14.4}{it}1}}}
\put(130,725){\makebox(0,0)[lb]{\smash{\SetFigFont{12}{14.4}{it}2}}}
\put( 25,825){\makebox(0,0)[lb]{\smash{\SetFigFont{12}{14.4}{it}2}}}
\end{picture}
\end{center}
The lines are the world-lines of the kinks, with their rapidities
assigned, while the spaces between the lines are indexed according to
the sectors between which the kinks interpolate.

One finds that to describe the amplitude of the above process the following 
substitutions have to be made in the S-matrix \cite{berlecl}: 
\begin{eqnarray}
{\hat {\cal R}} \rightarrow \sqrt{q}(-1)^{d+b-a-c}q^{C_a+C_c-C_b-C_d}
\left\{\matrix {\frac{1}{2} & a & b \cr \frac{1}{2} & c & d }\right\}_q
\ ,\nn
{\hat {\cal R}}^{-1} \rightarrow (\sqrt{q})^{-1}
(-1)^{d+b-a-c}q^{C_b+C_d-C_a-C_c}
\left\{\matrix {\frac{1}{2} & a & b \cr \frac{1}{2} & c & d }\right\}_q 
\ ,
\label{6jsubs}
\end{eqnarray}
where $C_a=a(a+1)$. 

The restriction procedure can be appropriately generalized to the case 
when the fundamental representation is spin-$1$ instead of spin-$1/2$ 
\cite{smirnov}. The
breather-soliton amplitudes and the breather-breather amplitudes
remain the same, since the breathers are singlets
(cf. \cite{smirresh}). It has been applied to the sine-Gordon models
to get S-matrices for $\Phi_{(1,3)}$-perturbed minimal models
\cite{smirresh,berlecl} and the
spin-$1$ generalization leads to the S-matrices of the
$\Phi_{(1,2)}$-perturbations using the ZMS model as starting point
\cite{smirnov, efthimiou}.

\subsection{The new RSOS restriction of the ZMS model}

Consider now the RSOS restriction of the ZMS model with respect
to the second quantum group ${\cal A}_0={\cal U}_{q^4}(sl(2))$. 
The fundamental
triplet splits into a doublet and a singlet under the action of 
${\cal A}_0$. This is where the subtlety of the procedure
lies. Since the representation is reducible, there are
possible mixing terms. The kink triplet splits into a doublet of what
can be called charged kinks, and a singlet neutral kink. Now one
may try to argue that since the neutral kink is a singlet, it has
nothing to do with the restriction procedure, like the
breathers. However, this state is degenerate in mass with the charged
kinks. Therefore there are amplitudes which describe the fusion of two
charged kinks into a neutral one and vice versa, which is essentially
different from the properties of the breathers. 

It is necessary to go over to a new gradation 
in which the rapidity dependence is
taken over by the generators in ${\cal A}_1$. It will be convenient to
use the variable
\begin{equation}
y=\sqrt{x}
\end{equation}
instead of $x$.
In this gradation, the quantum symmetry of the S-matrix is 
described by the following intertwining equations:
\begin{eqnarray}
&&{\tilde R}(y,q)\left( H_i\otimes 1+1 \otimes H_i \right)
=\left( H_i\otimes 1+1 \otimes H_i \right){\tilde R}(y,q) ,\nn
&&{\tilde R}(y,q)\left( E_1\otimes q^{-H_1/2}+yq^{H_1/2} 
\otimes E_1 \right) =
\left( E_1\otimes q^{H_1/2}+yq^{-H_1/2} \otimes E_1 \right)
{\tilde R}(y,q),\nn
&&{\tilde R}(y,q)\left( F_1\otimes q^{-H_1/2}
+\frac{1}{y}q^{H_1/2} \otimes F_1 \right) =
\left( F_1\otimes q^{H_1/2}
+\frac{1}{y}q^{-H_1/2} \otimes F_1 \right){\tilde R}(y,q),\nn
&&{\tilde R}(y,q)\left( E_0\otimes q^{-H_0/2}+q^{H_0/2} 
\otimes E_0 \right) =
\left( E_0\otimes q^{H_0/2}+q^{-H_0/2} \otimes E_0 \right)
{\tilde R}(y,q),\nn
&&{\tilde R}(y,q)\left( F_0\otimes q^{-H_0/2}+q^{H_0/2} 
\otimes F_0 \right) =
\left( F_0\otimes q^{H_0/2}+q^{-H_0/2} \otimes F_0 \right)
{\tilde R}(y,q). \nn
\end{eqnarray} 
The tensor part of the S-matrix is given by 
\begin{equation}
{\cal S}=P_{12}{\tilde R}(y,q)\ ,
\end{equation}
and has the following nonzero matrix elements:
\begin{eqnarray}
&& {\cal S}_{++}^{++}(y,q)=
 {\cal S}_{--}^{--}(y,q)=\frac{(y^2-q^4)(y^2+q^6)}{y^2q^5}\nn
&&{\cal S}_{+0}^{+0}(y,q)={\cal S}_{-0}^{-0}(y,q)=
{\cal S}_{0+}^{0+}(y,q)={\cal S}_{0-}^{0-}(y,q)=
-\frac{(q^4-1)(y^2+q^6)}{yq^5}\nn
&&{\cal S}_{+0}^{0+}(y,q)={\cal S}_{-0}^{0-}(y,q)=
{\cal S}_{0+}^{+0}(y,q)={\cal S}_{0-}^{-0}(y,q)=
\frac{(y^2+q^6)(y^2-1)}{y^2q^3}\nn
&&{\cal S}_{+-}^{-+}(y,q)={\cal S}_{-+}^{+-}(y,q)=
\frac{(y^2-1)(y^2+q^2)}{y^2q}\nn
&&{\cal S}_{+-}^{+-}(y,q)=-\frac{(q^4-1)(q^6+q^4y^2-q^4+y^2)}{q^5}\nn
&&{\cal S}_{-+}^{-+}(y,q)=-\frac{(q^4-1)(q^6+y^2-q^2+y^2q^2)}{q^5y^2}\nn
&&{\cal S}_{+-}^{00}={\cal S}_{00}^{+-}=
-q^4{\cal S}_{-+}^{00}=-q^4{\cal S}_{00}^{-+}=
-\frac{(q^4-1)(y^2-1)}{y}\nn
&&{\cal S}_{00}^{00}=
\frac{q^6y^2+y^2q^8-q^8-q^4y^2+y^2-q^{10}y^2+y^4q^2-y^2q^2}{y^2q^5}
\label{sregrad}\end{eqnarray}
This matrix ${\tilde R}(y,q)$ can, in fact, be obtained from the matrix
$R(x,q)$ in (\ref{hatS}) by a similarity transformation analogous to 
(\ref{chgrad}), hence it satisfies
the same unitarity relation and an appropriate crossing condition with a 
redefined $\alpha$ (see (\ref{crossuni})).

Proceeding to the RSOS restriction, let us first examine the charged
kink part. It is a $4\times 4$ submatrix of ${\cal S}$, which can be
expressed in terms of the fundamental R-matrix of 
${\cal U}_{q^4}(sl(2))$, analogously to the case of sine-Gordon
theory. This yields
\begin{equation}
\left(\frac{y^2}{q}-\frac{(1-q^2)q^3}{1+q^4}\right) 
{\hat{\cal R}}\left(q^4\right)^{-1}
-\left(\frac{q}{y^2}-\frac{1-q^2}{q(1+q^4)}\right) 
{\hat{\cal R}}\left(q^4\right)\ .
\end{equation}
Here
\begin{equation}
{\hat{\cal R}}\left(q^4\right)
\nonumber\end{equation}
denotes the matrix obtained from ${\hat{\cal R}}$, given in 
(\ref{doubletR}), by the substitution $q\rightarrow q^4$.

The prescription given in (\ref{6jsubs}) yields the following result,
using the explicit expression of the $6-j$ symbols given in 
the Appendix A of \cite{berlecl}:

\setlength{\unitlength}{0.012500in}%
\begingroup\makeatletter\ifx\SetFigFont\undefined
\def\x#1#2#3#4#5#6#7\relax{\def\x{#1#2#3#4#5#6}}%
\expandafter\x\fmtname xxxxxx\relax \def\y{splain}%
\ifx\x\y   
\gdef\SetFigFont#1#2#3{%
  \ifnum #1<17\tiny\else \ifnum #1<20\small\else
  \ifnum #1<24\normalsize\else \ifnum #1<29\large\else
  \ifnum #1<34\Large\else \ifnum #1<41\LARGE\else
     \huge\fi\fi\fi\fi\fi\fi
  \csname #3\endcsname}%
\else
\gdef\SetFigFont#1#2#3{\begingroup
  \count@#1\relax \ifnum 25<\count@\count@25\fi
  \def\x{\endgroup\@setsize\SetFigFont{#2pt}}%
  \expandafter\x
    \csname \romannumeral\the\count@ pt\expandafter\endcsname
    \csname @\romannumeral\the\count@ pt\endcsname
  \csname #3\endcsname}%
\fi
\fi\endgroup
\begin{picture}(55,45)(40,780)
\thinlines
\put( 20,780){\line( 1, 1){ 40}}
\put( 20,795){\makebox(0,0)[lb]{\smash{\SetFigFont{12}{14.4}{rm}a}}}
\put( 20,820){\line( 1,-1){ 40}}
\put( 55,795){\makebox(0,0)[lb]{\smash{\SetFigFont{12}{14.4}{rm}c}}}
\put( 75,795){\makebox(0,0)[lb]{\smash{\SetFigFont{12}{14.4}{it}= 
$\left(\displaystyle\frac{y^2}{q}-\frac{q}{y^2}-\frac{1}{q}+q\right)
\delta_{ac}
\left(\displaystyle \frac{[2b+1][2d+1]}{[2a+1][2c+1]}\right)^{1/2}+
\left(\displaystyle\frac{y^2}{q^5}-\frac{q^5}{y^2}-\frac{1}{q}+q\right)
\delta_{bd}$ }}}
\put( 35,780){\makebox(0,0)[lb]{\smash{\SetFigFont{12}{14.4}{rm}b}}}
\put( 35,810){\makebox(0,0)[lb]{\smash{\SetFigFont{12}{14.4}{rm}d}}}
\end{picture}
\begin{equation}
\end{equation}
for the charged kink scattering. The $q$-numbers in this equation are
defined with respect to $q^4$:
\begin{equation}
[x]=\frac{q^{4x}-q^{-4x}}{q^{4}-q^{-4}}\ .
\end{equation}
For the diagram to give a non-zero result,
neighbouring labels $a,b,c,d$ should differ by $\pm 1/2$. This
amplitude is not crossing symmetric, but needs a similarity
transformation to achieve crossing symmetry, like the original
S-matrix. The transformation is performed by multiplying the amplitude
with (c.f. \cite{berlecl})
\begin{equation}
\left(\displaystyle \frac{[2b+1][2d+1]}{[2a+1][2c+1]}\right)^{-\theta
/2\pi i}\ .
\end{equation}
The next step is to examine what happens to the neutral kink. 
Its presence means that it is possible to take two
neighbouring vacua to be the same. The following additional 
processes are allowed:
\begin{itemize}
\item Neutral kink scattering, neutral kink-charged kink forward
scattering and neutral kink-charged kink reflecion,
respectively. These are simple, since there are no
associated group structures.

\setlength{\unitlength}{0.012500in}%
\begingroup\makeatletter\ifx\SetFigFont\undefined
\def\x#1#2#3#4#5#6#7\relax{\def\x{#1#2#3#4#5#6}}%
\expandafter\x\fmtname xxxxxx\relax \def\y{splain}%
\ifx\x\y   
\gdef\SetFigFont#1#2#3{%
  \ifnum #1<17\tiny\else \ifnum #1<20\small\else
  \ifnum #1<24\normalsize\else \ifnum #1<29\large\else
  \ifnum #1<34\Large\else \ifnum #1<41\LARGE\else
     \huge\fi\fi\fi\fi\fi\fi
  \csname #3\endcsname}%
\else
\gdef\SetFigFont#1#2#3{\begingroup
  \count@#1\relax \ifnum 25<\count@\count@25\fi
  \def\x{\endgroup\@setsize\SetFigFont{#2pt}}%
  \expandafter\x
    \csname \romannumeral\the\count@ pt\expandafter\endcsname
    \csname @\romannumeral\the\count@ pt\endcsname
  \csname #3\endcsname}%
\fi
\fi\endgroup
\begin{equation}
\begin{picture}(370,45)(20,780)
\thinlines
\put( 20,780){\line( 1, 1){ 40}}
\put(160,820){\line( 1,-1){ 40}}
\put(160,780){\line( 1, 1){ 40}}
\put(300,780){\line( 1, 1){ 40}}
\put(300,820){\line( 1,-1){ 40}}
\put( 70,795){\makebox(0,0)[lb]{\smash{\SetFigFont{12}{14.4}{it}
$={\cal S}^{00}_{00}(\theta )$ }}}
\put(210,795){\makebox(0,0)[lb]{\smash{\SetFigFont{12}{14.4}{it}
$={\cal S}^{+0}_{0+}(\theta )$ }}}
\put(350,795){\makebox(0,0)[lb]{\smash{\SetFigFont{12}{14.4}{it}
$={\cal S}^{+0}_{+0}(\theta )$ }}}
\put( 40,810){\makebox(0,0)[lb]{\smash{\SetFigFont{12}{14.4}{rm}a}}}
\put( 20,820){\line( 1,-1){ 40}}
\put( 20,795){\makebox(0,0)[lb]{\smash{\SetFigFont{12}{14.4}{rm}a}}}
\put(180,810){\makebox(0,0)[lb]{\smash{\SetFigFont{12}{14.4}{rm}b}}}
\put( 55,795){\makebox(0,0)[lb]{\smash{\SetFigFont{12}{14.4}{rm}a}}}
\put(160,795){\makebox(0,0)[lb]{\smash{\SetFigFont{12}{14.4}{rm}a}}}
\put(195,795){\makebox(0,0)[lb]{\smash{\SetFigFont{12}{14.4}{rm}b}}}
\put(300,795){\makebox(0,0)[lb]{\smash{\SetFigFont{12}{14.4}{rm}a}}}
\put(320,810){\makebox(0,0)[lb]{\smash{\SetFigFont{12}{14.4}{rm}a}}}
\put(320,780){\makebox(0,0)[lb]{\smash{\SetFigFont{12}{14.4}{rm}a}}}
\put(335,795){\makebox(0,0)[lb]{\smash{\SetFigFont{12}{14.4}{rm}b}}}
\put( 40,780){\makebox(0,0)[lb]{\smash{\SetFigFont{12}{14.4}{rm}a}}}
\put(180,780){\makebox(0,0)[lb]{\smash{\SetFigFont{12}{14.4}{rm}a}}}
\end{picture}
\end{equation}

\item Two charged kinks turn into two neutral kinks. In this channel,
there is a Clebsh-Gordan coefficient to deal with. The presence of
this coefficient is responsible for the fact 
${\cal S}_{+-}^{00}=-q^4{\cal S}_{-+}^{00}$. This means that the
transition amplitude from the spin-$1$ state of the two charged kinks
to the two neutral kinks vanishes:
\begin{equation}
\frac{1}{q^2}{\cal S}_{+-}^{00}+q^2{\cal S}_{-+}^{00}=0.
\end{equation}
The amplitude for the process is given by the singlet component and
turns out to be
\setlength{\unitlength}{0.012500in}%
\begingroup\makeatletter\ifx\SetFigFont\undefined
\def\x#1#2#3#4#5#6#7\relax{\def\x{#1#2#3#4#5#6}}%
\expandafter\x\fmtname xxxxxx\relax \def\y{splain}%
\ifx\x\y   
\gdef\SetFigFont#1#2#3{%
  \ifnum #1<17\tiny\else \ifnum #1<20\small\else
  \ifnum #1<24\normalsize\else \ifnum #1<29\large\else
  \ifnum #1<34\Large\else \ifnum #1<41\LARGE\else
     \huge\fi\fi\fi\fi\fi\fi
  \csname #3\endcsname}%
\else
\gdef\SetFigFont#1#2#3{\begingroup
  \count@#1\relax \ifnum 25<\count@\count@25\fi
  \def\x{\endgroup\@setsize\SetFigFont{#2pt}}%
  \expandafter\x
    \csname \romannumeral\the\count@ pt\expandafter\endcsname
    \csname @\romannumeral\the\count@ pt\endcsname
  \csname #3\endcsname}%
\fi
\fi\endgroup
\begin{equation}
\begin{picture}(55,45)(40,780)
\thinlines
\put( 40,780){\line( 1, 1){ 40}}
\put( 40,795){\makebox(0,0)[lb]{\smash{\SetFigFont{12}{14.4}{rm}a}}}
\put( 40,820){\line( 1,-1){ 40}}
\put( 75,795){\makebox(0,0)[lb]{\smash{\SetFigFont{12}{14.4}{rm}a}}}
\put( 95,795){\makebox(0,0)[lb]{\smash{\SetFigFont{12}{14.4}{it}=
$\displaystyle i\frac{(q^4-1)(y^2-1)}{q^2y}$}}}
\put( 55,780){\makebox(0,0)[lb]{\smash{\SetFigFont{12}{14.4}{rm}b}}}
\put( 55,810){\makebox(0,0)[lb]{\smash{\SetFigFont{12}{14.4}{rm}a}}}
\end{picture}
\end{equation}
The factor $i$ is necessary to achieve crossing symmetry of the result.
(The crossing symmetry transformation in the variable $y$ takes the
form $y\rightarrow iq^3/y$, as can be seen from (\ref{crossuni})).
\item Two neutral kinks turn into two charged kinks. Since the
unreduced S-matrix is time-reflection symmetric, this amplitude is the
same as the one above.
\end{itemize}

This concludes the description of the reduced S-matrix. 

\section{Some explicit examples}\label{sect5}

\subsection{The model ${\cal M}_{(2,9)}+\Phi_{(1,5)}$}

The simplest models to consider are ${\cal M}_{(2,n)}$. 
In that case $q'^2=1$, which means that the maximal spin allowed is
$0$. Hence charged kinks are frozen. The models contain only the
neutral kink and the breathers. The related models are $\Phi_{(1,2)}$
perturbations of the models  ${\cal M}_{(8,n)}$. The particular case
of the S-matrix of the model ${\cal M}_{(2,9)}+\Phi_{(1,5)}$, which was 
derived and tested using the TBA and TCSA approach by Martins et al.
\cite{martins1, martins2}, 
can also be obtained in the framework presented here. Using 
the matrix element ${\cal S}_{00}^{00}$ in (\ref{sregrad}), after 
some computation the amplitude reduces to the form
\begin{equation}
-4i\sinh\frac{\pi}{\xi}\left(\theta + i\pi\right)
\sinh\frac{\pi}{\xi}\left(\theta + \frac{2i\pi}{3}\right)S_0(\theta )\ ,
\end{equation}
where $\xi=8\pi /15$.
From \cite{martins2} (formula (A.7)) one has 
\begin{eqnarray}
S_0(\theta )=
-\frac{1}{4i}\left( \sinh\frac{\pi}{\xi}\left(\theta + i\pi\right)
\sinh\frac{\pi}{\xi}\left(\theta + \frac{2i\pi}{3}\right)\right)^{-1}
\times \nn
f_{2/3}(\theta )f_{2/15}(\theta )f_{7/15}(\theta )
f_{-1/15}(\theta )f_{-2/5}(\theta )\ ,
\end{eqnarray}
with the notation
\begin{equation}
f_x(\theta )=\frac{\tanh\frac{1}{2}(\theta +ix\pi )}
{\tanh\frac{1}{2}(\theta -ix\pi )}\ .
\end{equation}
The final result for the amplitude is 
\begin{equation}
f_{2/3}(\theta )f_{2/15}(\theta )f_{7/15}(\theta )
f_{-1/15}(\theta )f_{-2/5}(\theta )\ ,
\end{equation}
which agrees with \cite{martins1,martins2}.

In \cite{martins2}, this amplitude was shown to correspond to a
particular combination of the amplitudes of 
${\cal M}_{(8,9)}+\Phi_{(1,2)}$, but the precise connection remained 
obscure. The present derivation shows that this special combination is
just the amplitude of the neutral kink in the unreduced theory, which
is (by virtue of the RSOS restriction framework) the S-matrix 
obtained after restriction.

More interesting for us are the models ${\cal M}_{(3,2n)}$, since 
${\cal M}_{(3,10)}$, ${\cal M}_{(3,14)}$ and ${\cal M}_{(3,16)}$, 
which are treated in detail in \cite{our}, fall into this class.

\subsection{The RSOS amplitudes for ${\cal M}_{(3,10)}$}

The minimal model ${\cal M}_{(3,10)}$ has central charge
$c=-44/5$. The Kac table consists of two rows and contains $9$
different conformal primary 
fields (which can be taken e.g. as the elements of the first
row), together with the identity. The field $\Phi_{(1,5)}$ has scaling
dimension $-2/5$, hence it generates a relevant perturbation of the
model. 

The S-matrix can be evaluated using the formulas in the previous
section. The related unitary model is ${\cal M}_{(5,6)}+\Phi_{(2,1)}$, 
which corresponds to
\begin{equation}
q=\exp\left(i\pi\frac{5}{6}\right)\ ,
\ q'=q^4=\exp\left(i\pi\frac{10}{3}\right)\ .
\end{equation}
Since $q'^3=1$, the maximum allowed spin is $1/2$. Hence charged kinks
are allowed and direct computation shows that there are only two
independent amplitudes. The eight amplitudes 
\setlength{\unitlength}{0.012500in}%
\begingroup\makeatletter\ifx\SetFigFont\undefined
\def\x#1#2#3#4#5#6#7\relax{\def\x{#1#2#3#4#5#6}}%
\expandafter\x\fmtname xxxxxx\relax \def\y{splain}%
\ifx\x\y   
\gdef\SetFigFont#1#2#3{%
  \ifnum #1<17\tiny\else \ifnum #1<20\small\else
  \ifnum #1<24\normalsize\else \ifnum #1<29\large\else
  \ifnum #1<34\Large\else \ifnum #1<41\LARGE\else
     \huge\fi\fi\fi\fi\fi\fi
  \csname #3\endcsname}%
\else
\gdef\SetFigFont#1#2#3{\begingroup
  \count@#1\relax \ifnum 25<\count@\count@25\fi
  \def\x{\endgroup\@setsize\SetFigFont{#2pt}}%
  \expandafter\x
    \csname \romannumeral\the\count@ pt\expandafter\endcsname
    \csname @\romannumeral\the\count@ pt\endcsname
  \csname #3\endcsname}%
\fi
\fi\endgroup
\begin{center}
\begin{picture}(285,115)(55,715)
\thinlines
\put( 60,780){\line( 1, 1){ 40}}
\put(140,820){\line( 1,-1){ 40}}
\put(140,780){\line( 1, 1){ 40}}
\put(220,820){\line( 1,-1){ 40}}
\put(220,780){\line( 1, 1){ 40}}
\put(300,820){\line( 1,-1){ 40}}
\put(300,780){\line( 1, 1){ 40}}
\put( 60,760){\line( 1,-1){ 40}}
\put( 60,720){\line( 1, 1){ 40}}
\put(140,760){\line( 1,-1){ 40}}
\put(140,720){\line( 1, 1){ 40}}
\put(220,760){\line( 1,-1){ 40}}
\put(220,720){\line( 1, 1){ 40}}
\put(300,760){\line( 1,-1){ 40}}
\put(300,720){\line( 1, 1){ 40}}
\put( 55,795){\makebox(0,0)[lb]{\smash{\SetFigFont{12}{14.4}{rm}0}}}
\put(100,795){\makebox(0,0)[lb]{\smash{\SetFigFont{12}{14.4}{rm}0}}}
\put( 78,815){\makebox(0,0)[lb]{\smash{\SetFigFont{12}{14.4}{rm}0}}}
\put( 78,775){\makebox(0,0)[lb]{\smash{\SetFigFont{12}{14.4}{rm}0}}}
\put( 70,755){\makebox(0,0)[lb]{\smash{\SetFigFont{12}{14.4}{rm}1/2}}}
\put( 55,735){\makebox(0,0)[lb]{\smash{\SetFigFont{12}{14.4}{rm}0}}}
\put( 90,735){\makebox(0,0)[lb]{\smash{\SetFigFont{12}{14.4}{rm}1/2}}}
\put( 78,715){\makebox(0,0)[lb]{\smash{\SetFigFont{12}{14.4}{rm}0}}}
\put( 60,820){\line( 1,-1){ 40}}
\put(135,735){\makebox(0,0)[lb]{\smash{\SetFigFont{12}{14.4}{rm}0}}}
\put(238,715){\makebox(0,0)[lb]{\smash{\SetFigFont{12}{14.4}{rm}0}}}
\put(150,715){\makebox(0,0)[lb]{\smash{\SetFigFont{12}{14.4}{rm}1/2}}}
\put(170,735){\makebox(0,0)[lb]{\smash{\SetFigFont{12}{14.4}{rm}1/2}}}
\put(158,755){\makebox(0,0)[lb]{\smash{\SetFigFont{12}{14.4}{rm}0}}}
\put(150,775){\makebox(0,0)[lb]{\smash{\SetFigFont{12}{14.4}{rm}1/2}}}
\put(135,795){\makebox(0,0)[lb]{\smash{\SetFigFont{12}{14.4}{rm}0}}}
\put(180,795){\makebox(0,0)[lb]{\smash{\SetFigFont{12}{14.4}{rm}0}}}
\put(150,815){\makebox(0,0)[lb]{\smash{\SetFigFont{12}{14.4}{rm}1/2}}}
\put(238,815){\makebox(0,0)[lb]{\smash{\SetFigFont{12}{14.4}{rm}0}}}
\put(210,795){\makebox(0,0)[lb]{\smash{\SetFigFont{12}{14.4}{rm}1/2}}}
\put(250,795){\makebox(0,0)[lb]{\smash{\SetFigFont{12}{14.4}{rm}1/2}}}
\put(238,775){\makebox(0,0)[lb]{\smash{\SetFigFont{12}{14.4}{rm}0}}}
\put(290,795){\makebox(0,0)[lb]{\smash{\SetFigFont{12}{14.4}{rm}1/2}}}
\put(310,775){\makebox(0,0)[lb]{\smash{\SetFigFont{12}{14.4}{rm}1/2}}}
\put(310,815){\makebox(0,0)[lb]{\smash{\SetFigFont{12}{14.4}{rm}1/2}}}
\put(330,795){\makebox(0,0)[lb]{\smash{\SetFigFont{12}{14.4}{rm}1/2}}}
\put(310,715){\makebox(0,0)[lb]{\smash{\SetFigFont{12}{14.4}{rm}1/2}}}
\put(290,735){\makebox(0,0)[lb]{\smash{\SetFigFont{12}{14.4}{rm}1/2}}}
\put(318,755){\makebox(0,0)[lb]{\smash{\SetFigFont{12}{14.4}{rm}0}}}
\put(340,735){\makebox(0,0)[lb]{\smash{\SetFigFont{12}{14.4}{rm}0}}}
\put(260,735){\makebox(0,0)[lb]{\smash{\SetFigFont{12}{14.4}{rm}0}}}
\put(210,735){\makebox(0,0)[lb]{\smash{\SetFigFont{12}{14.4}{rm}1/2}}}
\put(230,755){\makebox(0,0)[lb]{\smash{\SetFigFont{12}{14.4}{rm}1/2}}}
\end{picture}
\end{center}
are equal to
\begin{equation}
\frac{-i(y^2-1)^2}{y^2}\ ,
\label{ampl1}\end{equation}
while the other eight 
\setlength{\unitlength}{0.012500in}%
\begingroup\makeatletter\ifx\SetFigFont\undefined
\def\x#1#2#3#4#5#6#7\relax{\def\x{#1#2#3#4#5#6}}%
\expandafter\x\fmtname xxxxxx\relax \def\y{splain}%
\ifx\x\y   
\gdef\SetFigFont#1#2#3{%
  \ifnum #1<17\tiny\else \ifnum #1<20\small\else
  \ifnum #1<24\normalsize\else \ifnum #1<29\large\else
  \ifnum #1<34\Large\else \ifnum #1<41\LARGE\else
     \huge\fi\fi\fi\fi\fi\fi
  \csname #3\endcsname}%
\else
\gdef\SetFigFont#1#2#3{\begingroup
  \count@#1\relax \ifnum 25<\count@\count@25\fi
  \def\x{\endgroup\@setsize\SetFigFont{#2pt}}%
  \expandafter\x
    \csname \romannumeral\the\count@ pt\expandafter\endcsname
    \csname @\romannumeral\the\count@ pt\endcsname
  \csname #3\endcsname}%
\fi
\fi\endgroup
\begin{center}
\begin{picture}(285,115)(55,715)
\thinlines
\put( 60,780){\line( 1, 1){ 40}}
\put(140,820){\line( 1,-1){ 40}}
\put(220,820){\line( 1,-1){ 40}}
\put(220,780){\line( 1, 1){ 40}}
\put(300,780){\line( 1, 1){ 40}}
\put( 60,760){\line( 1,-1){ 40}}
\put( 60,720){\line( 1, 1){ 40}}
\put(140,760){\line( 1,-1){ 40}}
\put(140,720){\line( 1, 1){ 40}}
\put(220,760){\line( 1,-1){ 40}}
\put(220,720){\line( 1, 1){ 40}}
\put(300,760){\line( 1,-1){ 40}}
\put(300,720){\line( 1, 1){ 40}}
\put(140,780){\line( 1, 1){ 40}}
\put(300,820){\line( 1,-1){ 40}}
\put(135,735){\makebox(0,0)[lb]{\smash{\SetFigFont{12}{14.4}{rm}0}}}
\put( 55,795){\makebox(0,0)[lb]{\smash{\SetFigFont{12}{14.4}{rm}0}}}
\put( 70,815){\makebox(0,0)[lb]{\smash{\SetFigFont{12}{14.4}{rm}1/2}}}
\put(100,795){\makebox(0,0)[lb]{\smash{\SetFigFont{12}{14.4}{rm}0}}}
\put( 78,775){\makebox(0,0)[lb]{\smash{\SetFigFont{12}{14.4}{rm}0}}}
\put(135,795){\makebox(0,0)[lb]{\smash{\SetFigFont{12}{14.4}{rm}0}}}
\put(150,775){\makebox(0,0)[lb]{\smash{\SetFigFont{12}{14.4}{rm}1/2}}}
\put(152,755){\makebox(0,0)[lb]{\smash{\SetFigFont{12}{14.4}{rm}1/2}}}
\put( 60,820){\line( 1,-1){ 40}}
\put(160,815){\makebox(0,0)[lb]{\smash{\SetFigFont{12}{14.4}{rm}0}}}
\put(340,735){\makebox(0,0)[lb]{\smash{\SetFigFont{12}{14.4}{rm}0}}}
\put(180,795){\makebox(0,0)[lb]{\smash{\SetFigFont{12}{14.4}{rm}0}}}
\put(150,715){\makebox(0,0)[lb]{\smash{\SetFigFont{12}{14.4}{rm}1/2}}}
\put(175,735){\makebox(0,0)[lb]{\smash{\SetFigFont{12}{14.4}{rm}1/2}}}
\put( 95,735){\makebox(0,0)[lb]{\smash{\SetFigFont{12}{14.4}{rm}1/2}}}
\put( 55,735){\makebox(0,0)[lb]{\smash{\SetFigFont{12}{14.4}{rm}0}}}
\put( 78,715){\makebox(0,0)[lb]{\smash{\SetFigFont{12}{14.4}{rm}0}}}
\put( 78,755){\makebox(0,0)[lb]{\smash{\SetFigFont{12}{14.4}{rm}0}}}
\put(210,795){\makebox(0,0)[lb]{\smash{\SetFigFont{12}{14.4}{rm}1/2}}}
\put(228,815){\makebox(0,0)[lb]{\smash{\SetFigFont{12}{14.4}{rm}1/2}}}
\put(250,795){\makebox(0,0)[lb]{\smash{\SetFigFont{12}{14.4}{rm}1/2}}}
\put(238,775){\makebox(0,0)[lb]{\smash{\SetFigFont{12}{14.4}{rm}0}}}
\put(240,755){\makebox(0,0)[lb]{\smash{\SetFigFont{12}{14.4}{rm}0}}}
\put(210,735){\makebox(0,0)[lb]{\smash{\SetFigFont{12}{14.4}{rm}1/2}}}
\put(240,715){\makebox(0,0)[lb]{\smash{\SetFigFont{12}{14.4}{rm}0}}}
\put(260,735){\makebox(0,0)[lb]{\smash{\SetFigFont{12}{14.4}{rm}0}}}
\put(290,735){\makebox(0,0)[lb]{\smash{\SetFigFont{12}{14.4}{rm}1/2}}}
\put(310,755){\makebox(0,0)[lb]{\smash{\SetFigFont{12}{14.4}{rm}1/2}}}
\put(310,775){\makebox(0,0)[lb]{\smash{\SetFigFont{12}{14.4}{rm}1/2}}}
\put(290,795){\makebox(0,0)[lb]{\smash{\SetFigFont{12}{14.4}{rm}1/2}}}
\put(318,815){\makebox(0,0)[lb]{\smash{\SetFigFont{12}{14.4}{rm}0}}}
\put(330,795){\makebox(0,0)[lb]{\smash{\SetFigFont{12}{14.4}{rm}1/2}}}
\put(310,715){\makebox(0,0)[lb]{\smash{\SetFigFont{12}{14.4}{rm}1/2}}}
\end{picture}
\end{center}
are given by
\begin{equation}
\frac{\sqrt{3}(y^2-1)}{y}\ .
\label{ampl2}\end{equation}
For this model 
\begin{equation}
\xi=\pi\ ,\ 
y=\exp\left(\frac{\pi}{\xi}\theta\right)=\exp\left(\theta\right)\ ,
\end{equation}
where $\theta=\theta_1-\theta_2$ is the relative rapidity of the
particles. 
The amplitudes can be arranged into an $8\times 8$ matrix, with the
rows and columns indexed by sequences $\{ j_1,j_2,j_3\} ,\
j_i=0,1/2$. These sequences indicate the vacua between which the 
particles in the two-particle states mediate. The only possibly
non-zero matrix-elements are obtained only if the first and third
label of the incoming and outgoing two-particle states agrees, 
which means that the matrix is blockdiagonal with
$2\times 2$ blocks. In this form, it is easy to check the unitarity of
the transition amplitude.

\subsection{Particle interpretation}

The above S-matrix is in a ``kink picture'', which has the unusual
feature that not all possible sequences of kinks are allowed, only the
ones in which neighbouring kinks share the same vacuum states.   
Note that under the interchange of the vacua $0\leftrightarrow 1/2$,
which is a $\rm Z_2$ map, the amplitudes remain unchanged. Hence we
can choose to make an identification of the RSOS sequences in the
following way:  
\begin{equation}
\{j_1,j_2,\dots j_n\} \equiv \{1/2-j_1,1/2-j_2,\dots 1/2-j_n\}\ .
\end{equation}
Recalling that the  multiparticle state with $n-1$ particles
corresponding to an RSOS sequence $\{j_1,j_2,\dots j_n\}$ is given by:
\begin{equation}
K_{j_1j_2}(\theta_1)K_{j_2j_3}(\theta_2)\dots
K_{j_{n-1}j_n}(\theta_{n-1})\ , 
\end{equation} 
it is clear that the result is the identification $K_{0,0}\equiv
K_{1/2,1/2}$ and $K_{0,1/2}\equiv K_{1/2,0}$. Let us call $K$ 
the particle obtained from the first pair, and $\tilde K$ the 
one coming from the second. Then the scattering matrix reduces to a
four by four matrix describing the scattering of $K$ and ${\tilde
K}$. Their S-matrix reads as follows: 
\begin{eqnarray}
&& S^{KK}_{KK}(\theta )=
S^{{\tilde K}{\tilde K}}_{{\tilde K}{\tilde K}}(\theta )=
S^{{\tilde K}K}_{{\tilde K}K}(\theta )=
S^{K{\tilde K}}_{K{\tilde K}}(\theta )=
-4i\sinh^2(\theta )S_0(\theta )\nn 
&& S^{{\tilde K}{\tilde K}}_{KK}(\theta )=
S^{KK}_{{\tilde K}{\tilde K}}(\theta )=
S^{K{\tilde K}}_{{\tilde K}K}(\theta )
=S^{{\tilde K}K}_{K{\tilde K}}(\theta )=
4\sinh(\theta ) \sin\left(\frac{2\pi}{3} \right)S_0(\theta ) \nn
\end{eqnarray}
Using the identity 
\begin{equation}
\frac{\sinh\frac{\pi}{\zeta}\left(\theta
+i\alpha\frac{2}{3}\pi\right)}{\sinh\frac{\pi}{\zeta}\left(\theta
-i\alpha\frac{2}{3}\pi\right)}=
\exp\left( -i\int\limits_{-\infty}^\infty\frac{dk}{k}\sin k\theta 
\frac{\sinh\left(\frac{1}{2}\zeta-\alpha\frac{2}{3}\pi\right) k}
{\sinh\frac{1}{2}k\zeta}\right)
\end{equation}
and choosing the positive sign in (\ref{S0}),
the function $S_0(\theta )$ can be calculated with the result 
\begin{equation}
- \frac{1}{4i}\left(\sinh(\theta )\sinh\left(\theta 
-\frac{2\pi i}{3}\right)\right)^{-1}
\frac{\sinh\frac{1}{2}\left(\theta+\frac{\pi i}{3}\right)}
{\sinh\frac{1}{2}\left(\theta-\frac{\pi i}{3}\right)}\ .
\end{equation}
The eigenvalues of the two-particle S-matrix are  
\begin{equation}
1,\quad \left(\frac{1}{3}\right)\left(\frac{2}{3}\right)\ ,
\end{equation}
using the following notation common in the context of the 
S-matrix bootstrap
\begin{equation}
\left( p\right)=\frac{\sinh\left(\frac{\theta}{2}+p\frac{\pi i}{2}\right)}
{\sinh\left(\frac{\theta}{2}-p\frac{\pi i}{2}\right)}\ .
\end{equation} 
The two-particle states on which the scattering is diagonal, are just
symmetric and antisymmetric combinations of the two-particle states,
in terms of the particles $K$ and $\tilde K$: 
\begin{eqnarray}
|{K}(\theta_1){K}(\theta_2)\rangle\pm|
{\tilde K}(\theta_1){\tilde K}(\theta_2)\rangle\ ,
\ |K(\theta_1){\tilde K}(\theta_2)\rangle\pm
|{\tilde K}(\theta_1)K(\theta_2)\rangle\ .  
\end{eqnarray}
The eigenvalue $1$ corresponds to the antisymmetric, while the other
eigenvalue to the symmetric combinations. 
It is possible to define new one-particle states as follows:
\begin{equation}
|A\rangle = (|K\rangle + |{\tilde K}\rangle )/\sqrt{2}\ , \ 
|B\rangle = (|K\rangle - |{\tilde K}\rangle )/\sqrt{2}\ .
\end{equation}
We introduce a charge conjugation under which $A$ and $B$ are
selfconjugate particles. This is consistent, since the eigenvalues
above are transformed into themselves under crossing symmetry. 

The S-matrix in this basis is particularly simple, and the scattering
reduces to diagonal form:
\begin{equation}
S_{AA}=S_{BB}=\left(\frac{1}{3}\right)\left(\frac{2}{3}\right)\ ,\ 
S_{AB}=1\ .
\end{equation}
Note that $ S_{AA}$ and $S_{BB}$ are just two copies of  
the S-matrix of the model ${\cal M}_{(2,5)}+\Phi_{(1,2)}$, 
in which the spectrum consists of a self-conjugate scalar particle. 
This is not surprising if one notes that there is a different way of 
thinking about ${\cal M}_{(3,10)}$, namely,  as the tensor product 
${\cal M}_{(2,5)}\otimes{\cal M}_{(2,5)}$. The model 
${\cal M}_{(2,5)}$ has central charge $c=-22/5$ and contains 
two conformal families: one of them is given by the identity, 
the other one is generated by the operator $\phi_{(1,2)}$ 
with dimension $-2/5$.

${\cal M}_{(2,5)}\otimes{\cal M}_{(2,5)}$ contains two Virasoro
algebras given by the operators
\begin{equation}
1\otimes T(z)\ ,\ T(z)\otimes 1\ ,
\end{equation}
where $T(z)$ is the energy-momentum tensor in ${\cal M}_{(2,5)}$.
Their symmetric combination is the energy-momentum tensor in
${\cal M}_{(3,10)}$. 
The fields in ${\cal M}_{(3,10)}$ can be classified into 
$\rm Z_2$-even and $\rm Z_2$-odd fields with respect to the
$\rm Z_2$ map provided by flipping the tensor product. The fields
with Kac label $(1,n)$ in the ${\cal M}_{(3,10)}$
model give $\rm Z_2$-even fields when $n$ is odd and 
$\rm Z_2$-odd fields when $n$ is even. The two sectors are
the even and odd sector, respectively. The perturbing operator 
$\Phi_{(1,5)}$ is in the even sector and is nothing other than 
$(1\otimes\phi_{(1,2)} +\phi_{(1,2)} \otimes 1)/\sqrt{2}$. 
The above results clearly reflect this correspondence.

The S-matrix of ${\cal M}_{(2,5)}+\Phi_{(1,2)}$ has the $\phi^3$
property, i.e. the particle occurs as bound state of itself. This
property is equally valid at the level of the ${\cal M}_{(3,10)}$
model. The residue at the bound state pole has wrong sign, which 
reflects the nonunitarity of the theory.

\subsection{The model ${\cal M}_{(3,14)}+\Phi_{(1,5)}$}

Now let us turn to the model ${\cal M}_{(3,14)}+\Phi_{(1,5)}$. The
reduced amplitudes turn out to be identical to 
(\ref{ampl1},\ref{ampl2}), up to some changes in the sign. 
Now
\begin{equation}
\xi=\pi /2\ ,\ y=\exp(2\theta )\ . 
\end{equation}
The S-matrix takes the form
\begin{eqnarray}
&& S^{KK}_{KK}(\theta )=
S^{{\tilde K}{\tilde K}}_{{\tilde K}{\tilde K}}(\theta )=
S^{{\tilde K}K}_{{\tilde K}K}(\theta )=
S^{K{\tilde K}}_{{K}{\tilde K}}(\theta )=
4i\sinh^2(2\theta )S_0(\theta )\nn
&& -S^{{\tilde K}{\tilde K}}_{KK}(\theta )=
-S^{KK}_{{\tilde K}{\tilde K}}(\theta )=
S^{{\tilde K}K}_{K{\tilde K}}(\theta )
=S^{K{\tilde K}}_{{\tilde K}K}(\theta )=
4\sinh(2\theta ) \sin\left(\frac{2\pi}{3} \right)S_0(\theta )\nn
\label{signflip}\end{eqnarray}
Notice the sign flip in the second set of formulas. It will prove to
be important below.
In addition, $S_0$ now reads (choosing the negative sign in (\ref{S0}))
\begin{equation}
-\frac{1}{4i}\left(\sinh(2\theta )\sinh\left(2\theta 
-\frac{4\pi i}{3}\right)\right)^{-1}
\left(\frac{1}{3}\right)\left(\frac{1}{2}\right)
\left(\frac{5}{6}\right)\ .
\end{equation}
The eigenvalues of the two-particle transition amplitudes
turn out to be
\begin{equation}
-\left(\frac{1}{3}\right)\left(\frac{1}{2}\right)\left(\frac{1}{6}\right)
\ ,\ 
\left(\frac{2}{3}\right)\left(\frac{1}{2}\right)\left(\frac{5}{6}\right)\
,
\label{eig314}\end{equation}
however, now the first corresponds to the combinations
\begin{equation}
\ |K(\theta_1){\tilde K}(\theta_2) \rangle 
- |{\tilde K}(\theta_1)K(\theta_2)\rangle\ ,
|{K}(\theta_1){K}(\theta_2)\rangle
+ |{\tilde K}(\theta_1){\tilde K}(\theta_2)\rangle\ ,
\label{hkinkpoles}\end{equation}
while the second one to the states 
\begin{equation}
|{K}(\theta_1){K}(\theta_2)\rangle 
- |{\tilde K}(\theta_1){\tilde K}(\theta_2)\rangle\ ,
\ |K(\theta_1){\tilde K}(\theta_2) \rangle 
+ |{\tilde K}(\theta_1)K(\theta_2)\rangle\ .
\end{equation}
Due to the sign flip, each eigenvalue now corresponds to a symmetric
and an antisymmetric combination.
Note that the amplitudes in (\ref{eig314}) are just crossing symmetric
partners of each other. Therefore we are led to introduce the
following particles:
\begin{equation}
|A\rangle = (|K\rangle + i|{\tilde K}\rangle )/\sqrt{2}\ , \ 
|{\bar A}\rangle = (|K\rangle - i|{\tilde K}\rangle )/\sqrt{2}\ ,
\end{equation}
so that the S-matrix takes the form 
\begin{equation}
S_{AA}=S_{{\bar A}{\bar A}}= \left(\frac{2}{3}\right)
\left(\frac{1}{2}\right)\left(\frac{5}{6}\right)\ ,\ 
S_{A{\bar A}}=-\left(\frac{1}{3}\right)
\left(\frac{1}{2}\right)\left(\frac{1}{6}\right)\ ,
\end{equation}
and we can treat $A$ and ${\bar A}$ as conjugates of each other.

The unrestricted model contains a higher kink triplet and two
breathers as well. The kink-higher
kink and higher kink-higher kink S-matrices can be calculated 
using bootstrap for the bound state poles.

The higher kink pole is in the space spanned by the vectors
in (\ref{hkinkpoles}). We call $L$ the higher kink coming from the 
channel given by the second vector and $\tilde L$ the one coming 
from the first. Then the $KL$ S-matrix turns out to be the 
following:
\begin{eqnarray}
&& S^{KL}_{KL}(\theta )=
S^{{\tilde K}{\tilde L}}_{{\tilde K}{\tilde L}}(\theta )=
S^{{\tilde K}L}_{{\tilde K}L}(\theta )=
S^{K{\tilde L}}_{{K}{\tilde L}}(\theta )=\nn
&&\frac{2(-1+i\sqrt{3})(y^2+1)}
{(2y+\sqrt{3}+i)(2y-\sqrt{3}-i)}
S_0^{KL}(\theta )\nn
&& -S^{{\tilde K}{\tilde L}}_{KL}(\theta )=
-S^{KL}_{{\tilde K}{\tilde L}}(\theta )=
S^{{\tilde K}L}_{K{\tilde L}}(\theta )
=S^{K{\tilde K}}_{{\tilde K}K}(\theta )=\nn
&&-\frac{2(-1+i\sqrt{3})\sqrt{3}y}
{(2y+\sqrt{3}+i)(2y-\sqrt{3}-i)}
S_0^{KL}(\theta )\ ,
\label{KLscatt}\end{eqnarray}
where
\begin{equation}
S_0^{KL}(\theta )=
\left(\frac{1}{4}\right)\left(\frac{3}{4}\right)
\left(\frac{5}{12}\right)^2\left(\frac{7}{12}\right)
\left(\frac{11}{12}\right)\ .
\end{equation}
The S-matrix $S^{KL}(\theta )$ given in (\ref{KLscatt}) is pseudounitary, 
i.e. satisfies
\begin{equation}
S^{KL}(-\theta )AS^{KL}(\theta )A=I\ ,
\end{equation}
where $I$ is the $4\times 4$ unit matrix and $A={\rm diag}(1,1,-1,-1)$.
Diagonalizing $S^{KL}(\theta )$ we find the eigenvalues
\begin{eqnarray}
S^{KL}_1(\theta )=\left(\frac{1}{4}\right)\left(\frac{3}{4}\right)
\left(\frac{5}{12}\right)\left(\frac{11}{12}\right)
\frac{\sinh\frac{1}{2}\left(\theta+\frac{5}{12}i\pi\right)
\sinh\frac{1}{2}\left(\theta+\frac{7}{12}i\pi\right)}
{\sinh\frac{1}{2}\left(\theta-\frac{1}{12}i\pi\right)
\sinh\frac{1}{2}\left(\theta-\frac{11}{12}i\pi\right)}
\ ,\nn 
S^{KL}_2(\theta )=\left(\frac{1}{4}\right)\left(\frac{3}{4}\right)
\left(\frac{5}{12}\right)\left(\frac{11}{12}\right)
\frac{\sinh\frac{1}{2}\left(\theta+\frac{1}{12}i\pi\right)
\sinh\frac{1}{2}\left(\theta+\frac{11}{12}i\pi\right)}
{\sinh\frac{1}{2}\left(\theta-\frac{5}{12}i\pi\right)
\sinh\frac{1}{2}\left(\theta-\frac{7}{12}i\pi\right)}\ .
\end{eqnarray}
The eigenvectors look very similar as in the case of the fundamental 
kink S-matrix. In the basis of the eigenvectors $A$ takes the form
\begin{equation}
A=\left(\matrix{0&0&0&1\cr 0&0&1&0\cr 0&1&0&0\cr 1&0&0&0}\right)\ .
\end{equation}
So pseudounitarity means that the eigenvalues are not pure phases but 
rather satisfy
\begin{equation}
S^{KL}_1(\theta )S^{KL}_2(-\theta )=1\ .
\end{equation}
The appearance of such matrices $A$ after RSOS restriction was already 
noticed in \cite{smirresh}. They are allowed by quantum group symmetry 
since it only tells us that states corresponding to different RSOS 
sequences must be orthogonal (cf. the discussion in section 4.1 and 
references therein), hence $A$ should be diagonal in the basis 
of RSOS states, with its diagonal entries being $\pm 1$ when properly 
normalized. $A$ can be thought of as a metric on the 
state space (in this case the subspace of kink-higher kink two-particle 
states) and is (partially at least) fixed by the pseudounitarity requirement. 
It must coincide with the metric on the state space originating from 
the unperturbed CFT. Note that the eigenvectors of $S^{KL}(\theta )$ 
are of zero pseudonorm with respect to $A$ and also that the eigenvalues 
could have been obtained by applying diagonal bootstrap rules starting
from the fundamental kink phaseshifts.

From (\ref{higherkinkS}) we learn that the higher kink-higher kink
S-matrix is proportional to the fundamental kink S-matrix. Therefore
one can introduce the states $B$ and ${\bar B}$, following the analogy
of $A$ and ${\bar A}$. These states will then diagonalize the kink
S-matrices. The remaining S-matrices present no additional novelties 
and are the following:
\begin{eqnarray}
&&S_{BB}=S_{{\bar B}{\bar B}}=
\left(\frac{1}{6}\right)\left(\frac{5}{6}\right)^2
\left(\frac{1}{2}\right)^3\left(\frac{1}{3}\right)^2
\left(\frac{2}{3}\right)^3
\ ,\nn 
&&S_{B{\bar B}}=
-\left(\frac{1}{6}\right)^2\left(\frac{5}{6}\right)
\left(\frac{1}{2}\right)^3\left(\frac{1}{3}\right)^3
\left(\frac{2}{3}\right)^2\ ,\nn 
&&S_{AC}=S_{{\bar A}C}=
\left(\frac{1}{4}\right)\left(\frac{3}{4}\right)
\left(\frac{5}{12}\right)\left(\frac{7}{12}\right)\ ,\nn  
&&S_{BC}=S_{{\bar B}C}=\left(\frac{1}{6}\right)
\left(\frac{5}{6}\right)\left(\frac{1}{2}\right)^2
\left(\frac{1}{3}\right)^2\left(\frac{2}{3}\right)^2\ ,\nn
&&S_{CC}=
-\left(\frac{1}{3}\right)\left(\frac{2}{3}\right)
\left(\frac{1}{6}\right)\left(\frac{5}{6}\right)
\left(\frac{1}{2}\right)^2\ ,\nn
&&S_{AD}=S_{{\bar A}D}=\left(\frac{1}{6}\right)
\left(\frac{5}{6}\right)\left(\frac{1}{3}\right)^2
\left(\frac{2}{3}\right)^2\left(\frac{1}{2}\right)^2\ ,\nn
&&S_{BD}=S_{{\bar B}D}=\left(\frac{1}{12}\right)
\left(\frac{11}{12}\right)\left(\frac{1}{4}\right)^3
\left(\frac{3}{4}\right)^3\left(\frac{5}{12}\right)^4
\left(\frac{7}{12}\right)^4\ ,\nn
&&S_{CD}=\left(\frac{1}{12}\right)\left(\frac{11}{12}\right)
\left(\frac{1}{4}\right)^2\left(\frac{3}{4}\right)^2
\left(\frac{5}{12}\right)^3\left(\frac{7}{12}\right)^3\ ,\nn
&&S_{DD}=-\left(\frac{1}{6}\right)^3\left(\frac{5}{6}\right)^3
\left(\frac{1}{3}\right)^5\left(\frac{2}{3}\right)^5
\left(\frac{1}{2}\right)^6\ .
\end{eqnarray} 
The masses in the model are given by
\begin{eqnarray}
m_A=m_{\bar A}=m,\ m_B=m_{\bar B}=2m\cos\left(\frac{\pi}{12}\right),\nn
m_C=2m\cos\left(\frac{\pi}{4}\right),\ 
m_D=4m\cos\left(\frac{\pi}{12}\right)\cos\left(\frac{\pi}{4}\right)\ . 
\end{eqnarray}
The first is the fundamental kink mass, the next one is the mass of
the higher kink and the last two are the two breathers. 

The S-matrix has a $\rm Z_2$ invariance, which exchanges $A$ with
$\bar A$ and $B$ with $\bar B$, while it leaves $C$ and $D$
invariant. This is to be expected from the fact that the minimal model
${\cal M}_{(3,14)}$ (similarly to ${\cal M}_{(3,10)}$) has a $\rm Z_2$
invariance and the perturbing operator is $\rm Z_2$-even.

To establish consistency of the above picture, it is also useful 
to check the bootstrap consistency equations which constrain the 
higher spin conserved charges. For the $E_6$ case of the related 
unitary theory ${\cal M}_{(6,7)}+\Phi_{(1,2)}$, 
this has been done in \cite{E6}. The consistency equations given there 
can be easily changed to reflect the new fusion rules and they still 
allow for charges with spins 
\begin{equation}
s\equiv 1\ , 5 \ {\rm mod}\ 6\ .
\end{equation}
These charges  
are $\rm Z_2$-invariant, i.e., they take the same value on $A$ and 
$\bar A$ and similarly for $B$ and $\bar B$. This is just the 
$\rm Z_2$-even subset of the charges allowed by the $E_6$ fusion rules.
A further check of the above S-matrix is provided 
by the TCSA and TBA analysis given in \cite{our}.

Let me now make some remarks on the sign flip, observed in 
(\ref{signflip}), which is characteristic for all 
${\cal M}_{(3,4n+2)}$ models. The reason is that that the signs depend 
on the arithmetic properties of $q$. Generically, the models of the
form ${\cal M}_{(3,4n+2)}$ are diagonalizable in terms of
self-conjugate particles, if 
\begin{equation}
4n+2\equiv 10,26 \ {\rm mod}\ 24  
\end{equation}
(since the tensor
part of the amplitude is the same as for ${\cal M}_{(3,10)}$, the
only difference is in the definition of $y$ in terms of $\theta$ and
in the factor $S_0(\theta )$), and have flipped assignment of
eigenvectors and therefore diagonalizable in terms of conjugate
particle pairs if 
\begin{equation}
4n+2\equiv 14,22 \ {\rm mod}\ 24 \ , 
\end{equation}
as can be proven by straightforward calculation of the amplitudes. 
The period $24$ just
reflects the periodic dependence of $q$ on $n$. I would like to remark 
that even though the S-matrix is diagonalizable, the two particles are 
not independent, except in the case ${\cal M}_{(3,10)}$, since the 
amplitude $S_{AB}$ is generically different from $1$. The 
${\cal M}_{(3,10)}$ case is special due to its tensor product form 
in terms of ${\cal M}_{(2,5)}$.  

\subsection{The model ${\cal M}_{(3,16)}+\Phi_{(1,5)}$}

Let us close this section with the model 
${\cal M}_{(3,16)}+\Phi_{(1,5)}$. The corresponding unitary model is 
${\cal M}_{(3,4)}+\Phi_{(1,2)}$, the magnetic perturbation of the Ising
model. In the Ising case $q=\exp(4i\pi /3)$, hence the allowed maximal spin
is $1/2$ and since the kinks are in the triplet
representation of ${\cal A}_1$,
all the kink degrees of freedom are frozen and only
breathers remain in the spectrum. There are 8 breathers and the
scattering is described by the so-called $E_8$ S-matrix \cite{smirnov}.

However, in the new restriction, the kinks have singlet components so
they can never be frozen. 
In addition, since $q'=\exp(16\pi i/3)$ gives $j_{max}=1/2$, the
charged kinks are allowed to remain in the spectrum as well. 
Therefore one can expect to see
the masses of the kink and the higher kinks in the spectrum. This is
different from the models ${\cal M}_{(3,10)}+\Phi_{(1,5)}$ and 
${\cal M}_{(3,14)}+\Phi_{(1,5)}$, where the spectra (superficially at
least) are the same as those of the corresponding $\Phi_{(1,2)}$ 
perturbed CFT. The
term superficially indicates that the scattering of the particles is
different in ${\cal M}_{(3,10)}+\Phi_{(1,5)}$ and 
${\cal M}_{(3,14)}+\Phi_{(1,5)}$ from that of the related unitary
models. 

The S-matrix can be written in a very similar form as in the 
${\cal M}_{(3,10)}+\Phi_{(1,5)}$ and ${\cal M}_{(3,14)}+\Phi_{(1,5)}$
case. Explicitly, it is given by:
\begin{eqnarray}
&&S_{KK}^{KK}=\frac{y^4-2i\sqrt{3}y^2-1}{y^2}S_0(\theta )\ ,\ 
S_{{\tilde K}{\tilde K}}^{{\tilde K}{\tilde K}}=
-\frac{y^4+2i\sqrt{3}y^2-1}{y^2}S_0(\theta )\ ,\nn 
&&S_{{\tilde K}{K}}^{{K}{\tilde K}}
=S_{{K}{\tilde K}}^{{\tilde K}{K}}=
-i\sqrt{3}\frac{y^2+1}{y}S_0(\theta )\ , \ 
S_{{\tilde K}{\tilde K}}^{{K}{K}}=
S_{{K}{K}}^{{\tilde K}{\tilde K}}=
-\sqrt{3}\frac{y^2-1}{y}S_0(\theta )\ ,\nn
&&S_{{\tilde K}{K}}^{{\tilde K}{K}}=
S_{{K}{\tilde K}}^{{K}{\tilde K}}=
\frac{y^4-1}{y^2}S_0(\theta )\ , \ 
\end{eqnarray} 
where
\begin{equation}
y=\exp\left(\frac{5\theta}{2}\right)\ ,
\end{equation}
and (choosing the positive sign in (\ref{S0}))
\begin{eqnarray}
S_0(\theta )=\frac{1}{4i}\left( \sinh\frac{5}{2}\left(\theta-\pi i\right)
\sinh\frac{5}{2}\left(\theta-\frac{2\pi i}{3}\right)\right)^{-1}\times\nn
\exp\left( -2i\int\limits_0^\infty 
\frac{\sin k\theta \sinh \frac{\pi k}{3} \cosh\frac{\pi}{30} k}
{k\cosh\frac{\pi k}{2}\sinh\frac{\pi k}{5} } dk \right)\ ,
\label{S0316}\end{eqnarray}
since $\xi =2\pi /5$.

However, the integral (\ref{S0316}) cannot be carried out in closed 
form. In addition, it turns out, that while part
of the S-matrix is diagonalizable on rapidity-independent combinations
of the RSOS states, there is a part, which is not. 
Direct computations show that this is the case in all 
${\cal M}_{(3,4n)}$ models. This phenomenon can be retraced to the
fact that the arithmetical properties of $q$ are different in the two
class of models given by ${\cal M}_{(3,4n+2)}$ and ${\cal M}_{(3,4n)}$.

Let me spell out the eigenvalues, which is useful for the TCSA check
\cite{our}.
The first two eigenvalues of the two-particle S-matrix are:
\begin{eqnarray}
\frac{\sinh\left(\frac{5}{4}\theta -i\frac{\pi}{6}\right)}
{\sinh\left(\frac{5}{4}\theta +i\frac{\pi}{6}\right)}
\exp\left( -2i\int\limits_0^\infty \frac{\sin k\theta \sinh \frac{\pi
k}{3} \cosh\frac{\pi}{30} k}{k\cosh\frac{\pi
k}{2}\sinh\frac{\pi k}{5} } dk \right)\ ,\nn
\frac{\sinh\left(\frac{5}{4}\theta +i\frac{\pi}{3}\right)}
{\sinh\left(\frac{5}{4}\theta -i\frac{\pi}{3}\right)}
\exp\left( -2i\int\limits_0^\infty \frac{\sin k\theta \sinh \frac{\pi
k}{3} \cosh\frac{\pi}{30} k}{k\cosh\frac{\pi
k}{2}\sinh\frac{\pi k}{5} } dk \right)\ ,
\end{eqnarray}
corresponding to the vectors
\begin{equation}
\ |K(\theta_1){\tilde K}(\theta_2)\rangle 
+|{\tilde K}(\theta_1)K(\theta_2)\rangle\ , 
\ |K(\theta_1){\tilde K}(\theta_2)\rangle 
- |{\tilde K}(\theta_1)K(\theta_2)\rangle\ , 
\end{equation}
respectively, 
while the second pair takes the more complicated form
\begin{eqnarray}
\frac{1}{4}
\frac{2i\sqrt{3}+2\sinh\left(\frac{5}{2}\theta\right) 
\sqrt{2\cosh\left(5\theta\right) + 5}}
{\cosh\frac{5}{2}\theta
\sinh\left(\frac{5}{2}\theta-\frac{2\pi i}{3}\right)}
\exp\left(-2i\int\dots\right)\ ,\nn
\frac{1}{4}
\frac{2i\sqrt{3}-2\sinh\left(\frac{5}{2}\theta\right) 
\sqrt{2\cosh\left(5\theta\right) + 5}}
{\cosh\frac{5}{2}\theta
\sinh\left(\frac{5}{2}\theta-\frac{2\pi i}{3}\right) }
\exp\left(-2i\int\dots\right)\ ,
\end{eqnarray}
(the dots denote the same integrand as above) 
and corresponds to the vectors
\begin{eqnarray}
|K(\theta_1)K(\theta_2)\rangle 
+\frac{1}{\sqrt{3}}\left(2\cosh\frac{5}{2}\theta 
+\sqrt{2\cosh5\theta +5}\right)
|{\tilde K}(\theta_1){\tilde K}(\theta_2)\rangle\ ,\nn
|K(\theta_1)K(\theta_2)\rangle 
+\frac{1}{\sqrt{3}}\left(2\cosh\frac{5}{2}\theta 
-\sqrt{2\cosh5\theta +5}\right)
|{\tilde K}(\theta_1){\tilde K}(\theta_2)\rangle\ ,
\end{eqnarray}
with $\theta=\theta_1-\theta_2$ 
being the rapidity difference between the particles.

Due to the fact that the eigenvalues do not come in doubly degenerate 
pairs, in contrast to the case of the models ${\cal M}_{(3,4n+2)}$, 
there are no one-particle states on which the scattering can be 
diagonalized.

\section{Conclusions}\label{sect6}

The results described above show 
that due to fact that the algebra $A_2^{(2)}$ is
non-simply laced, the ZMS model allows another RSOS restriction,
different from the one previuosly known. The S-matrices and the
spectra of these theories can be derived by traditional methods of
exact S-matrix theory and using the RSOS restriction procedure. 
The new restriction corresponds to the $\Phi_{(1,5)}$ perturbation 
of a different minimal model. The models ${\cal M}_{(3,10)}$, 
${\cal M}_{(3,14)}$ and ${\cal M}_{(3,16)}$ 
have been investigated in detail. It is clear from the examples that
the new RSOS restrictions have different S-matrices than the original 
$\Phi_{(1,2)}$ theory. In the  ${\cal M}_{(3,16)}+\Phi_{(1,5)}$ case, 
even the mass spectrum is different. Therefore, even if one
starts with the same unrestricted ZMS model, the two possible 
restrictions yield completely different physics.

As mentioned in the introduction, it would be of interest to extend
the results of this paper to imaginary coupling affine Toda field
theories based on more general non-simply laced affine Kac-Moody
algebras, and to learn more about how the way, in which the RSOS
restriction is performed, influences the physical picture obtained 
after the restriction. 

The results presented here have been checked by applying the truncated
conformal space approach combined with thermodynamical Bethe
Ansatz calculations \cite{our}. The results of the TCSA and TBA
calculations are in complete accord with the details of the
S-matrices and spectra described in this paper. 
\vspace{.5in}
\begin{center}
{\bf Acknowledgements}
\end{center}

I would like to thank Dr. G.~M.~T.~Watts for directing my attention to
the problem and for the many useful discussions we had in the
course of this work. I am very grateful to the Cambridge Overseas
Trust for providing the funds for my visit in Cambridge, and the
Department of Applied Mathematics and Theoretical Physics for their
warm hospitality. Some of the computations were performed using 
computers purchased on EPSRC (UK) grant GR/J73322. I would also like 
to thank Dr. M. J. Martins for bringing the papers
\cite{martins1,martins2} to my attention and Prof. E. Corrigan for 
his very useful comments.

\end{document}